\documentclass[a4paper,12pt]{article}
\usepackage{jheppub,booktabs}
\usepackage{amsmath,amssymb}
\usepackage{xcolor,colortbl}
\usepackage{cancel}
\usepackage{subfigure}
\usepackage{mathrsfs}
\usepackage{graphicx}
\usepackage{hepunits}
\usepackage{wasysym}
\usepackage{feynmf}
\usepackage{enumerate}
\usepackage{tikz}
\usepackage[utf8x]{inputenc}
\usetikzlibrary{fit}

\tikzset{   
        every picture/.style={remember picture,baseline},
        every node/.style={anchor=base,align=center,outer sep=1.5pt},
        every path/.style={thick},
        }

\newcommand\marktopleft[1]{%
    \tikz[overlay,remember picture] 
        \node (marker-#1-a) at (-.0em,.3em) {};%
}
\newcommand\markbottomright[1]{%
    \tikz[overlay,remember picture] 
        \node (marker-#1-b) at (-.3em,.33em) {};%
    \tikz[overlay,remember picture,inner sep=3pt]
        \node[draw=orange,dashed,rounded corners,fit=(marker-#1-a.north west) (marker-#1-b.south east)] {};%
}

\newcommand\markbottomrightblue[1]{%
    \tikz[overlay,remember picture] 
        \node (marker-#1-b) at (4.8em,.33em) {};%
    \tikz[overlay,remember picture,inner sep=3pt]
        \node[draw=blue,dashed,rounded corners,fit=(marker-#1-a.north west) (marker-#1-b.south east)] {};%
}

\author{Florian Goertz}
\affiliation{Max-Planck-Institut f{\"u}r Kernphysik\\ Saupfercheckweg 1, 69117 Heidelberg, Germany,}
\affiliation{\vspace*{2mm}
Theory Division,\\
CERN, 1211 Geneva 23, Switzerland}

\emailAdd{fgoertz@mpi-hd.mpg.de}

\title{Accessing Masses Beyond Collider Reach - in EFT}

\abstract{We demonstrate how masses of new states, beyond direct experimental 
reach, could nevertheless be extracted in the framework of effective field theory (EFT),
given broad assumptions on the underlying UV physics, however not sticking to a particular 
setup nor fixing the coupling strength of the scenario. The flat direction in the $g_\ast$
vs. $M$ plane is lifted by studying correlations between observables that depend on 
operators with a different $\hbar$ scaling. We discuss the remaining model dependence 
(which is inherent even in the EFT approach to have control over the error 
due to the truncation of the power series), as well as prospects to test paradigms of UV physics.
In particular, we provide an assessment of which correlations are best suited regarding  
sensitivity, give an overview of possible/expected effects in different observables,
and demonstrate how perturbativity and direct search limits corner possible 
patterns of deviations from the SM in a given UV paradigm.}
\date{\today}

\begin{document}
\maketitle

\section{General Introduction}

Given the clear technical limitations in increasing the energy of collider experiments,
it is important to try to access the properties of physics beyond the Standard Model (BSM)
indirectly via precise measurements of observables at available energies. However,
this approach features in general a significant limitation, since new degrees of freedom,
entering as virtual particles in processes with characteristic energy scales below their production 
threshold, generically lead to corrections scaling like a ratio of their coupling and their mass.
Acquiring information about their actual mass spectrum requires usually very specific assumptions
about the coupling strength - which in reality however could span a huge range from
very weak coupling $g_\ast \ll 1$ up to strong coupling, like $g_\ast \lesssim 4 \pi$.
In this paper we point out how the different $\hbar$ scaling of various operators can be used to lift 
flat directions in the coupling vs. mass plane by examining more than one observable at a time,
thus allowing to estimate systematically at which mass the new physics (NP) actually 
can be expected.

The framework used for this analysis is the effective field theory (EFT) extension of the SM, which
is indeed the most general parametrization of NP that resides at energies
much larger than both the electroweak scale $M_{\rm EW}$ and the characteristic scale $E$ probed by the experiment 
of interest. It can be fully formulated in terms of low mass (SM) fields, while the effect of the NP will 
manifest itself in the presence of operators with mass dimension $D>4$ in the effective Lagrangian \cite{Buchmuller:1985jz,Hagiwara:1993ck,Grzadkowski:2010es}
\begin{equation}
{\cal L}_{\rm{eff}}={\cal L}_{\rm SM}+ \sum_{i}  c_{i}^{(6)} {\cal O}_{i}^{(6)} \
+ \sum_{j}  c_{j}^{(8)} {\cal O}_{j}^{(8)}  + \cdots\,,
\end{equation}
where ${\cal L}_{\rm SM}$ is the SM Lagrangian, and we have assumed baryon and lepton number conservation. 
The operators ${\cal O}_{i}^{(D)}$ of canonical dimension $D$ consist of all (Poincar{\'e} invariant and hermitian) combinations
of SM fields that respect local invariance under the (linearly-realized) $SU(3)_c \times SU(2)_L \times U(1)_Y$
SM symmetry group. On dimensional grounds, operators of higher $D$ will be suppressed by larger powers of some 
fundamental mass scale $M \gg E$, associated to the new states that have been removed as propagating 
degrees of freedom in the low energy theory ${\cal L}_{\rm{eff}}$ (i.e. integrated out in the path integral), 
with their effect being contained in the coefficients $c_{i}^{(6)}$ \cite{Weinberg:1966fm,Weinberg:1968de,Coleman:1969sm,Callan:1969sn,Dashen:1969eg,Dashen:1969ez,Li:1971vr,Wilson:1971,Wilson:1973jj,Appelquist:1974tg,Weinberg:1978kz,Weinberg:1979sa,Wilczek:1979hc,Weldon:1980gi,Weinberg:1980wa} (for reviews and further developments see, {\it e.g.,} \cite{Polchinski:1992ed,Georgi:1994qn,Pich:1998xt,Buras:1998raa,Neubert:2005mu,Burgess:2007pt,Falkowski:2015fla,deFlorian:2016spz}).\footnote{This suppression allows for a truncation of the series at 
a certain $D$, assuring the predictivity of the setup.} These coefficients allow 
to capture the effect of NP, no matter what are the exact details of the theory at higher energies. 
Determining them in experiment is a first step to understand the UV completion of the SM. On the other 
hand, as mentioned above, they generically depend only on {\it ratios} of couplings over masses, and not 
on masses alone, so naively it seems not possible to fix the {\it spectrum} of NP from such low energy observations,
such as to know where to search for it.\footnote{See also \cite{Giudice:2016yja} for a recent 
discussion on the difference of new mass thresholds and (combined) interaction scales.}

However, one needs to realize that not all operators depend on the very {\it same} ratio.
In fact, from restoring $\hbar$ dimensions in the action and simple dimensional analysis, 
it is easy to convince oneself, that very generally an operator containing $n_i$ fields 
features a coefficient scaling as
\begin{equation}
c^{(D)}_i \sim \frac{(\text{coupling})^{n_i-2}}{(\text{high mass scale})^{D-4}}\, ,
\end{equation}
given that the UV theory is perturbative (see, e.g.,~\cite{Luty:1997fk,Cohen:1997rt,Giudice:2007fh}).\footnote{Note 
that an additional suppressing factor $(\text{coupling}/4\pi)^{2L}$ can arise if the operator is 
only generated at the $L^{\rm th}$-loop order.}
Analyzing the effect of more than one operator at a time can provide information about $M$.

In fact, as we will study in detail below, exploring operators with a different field content allows us 
to gain sensitivity on different ratios of coupling over mass such as to solve for the latter.
The broad assumptions required to entertain such correlations will be detailed in the following section.
They correspond to a set of power counting rules, which are required in any case to assess the validity
of the EFT setup, 
and do not include specific assumptions, such as on concrete coupling strengths.
This article is arganized as follows. 
In Section~\ref{sec:setup}, we will provide more details on the setup that forms the basis for our analysis,
in Section~\ref{sec:ana}, we will perform the actual simultaneous study of different observables that
will allow us to learn something on the underlying mass scale and to test UV paradigms, 
while we will conclude in Section~\ref{sec:conc}.

\section{Setup}
\label{sec:setup}
In the following, we will detail the power counting rules used in the analysis at hand. 
They basically correspond to the assumption of one new scale $M$ and one NP coupling $g_\ast$, in the 
spirit of the Strongly-Interacting Light Higgs (SILH) \cite{Giudice:2007fh}, and are fulfilled in a broad 
class of models where a weakly coupled narrow resonance (characterized by a single coupling constant) is integrated out 
but as well in strongly coupled NP setups featuring a large $N$ description. Note that the assumption of power counting rules is crucial in any 
EFT if one wants to assess its validity, since only after fixing the scaling of operators with couplings and masses 
can one start to estimate the effect of truncating the EFT series at a certain mass dimension (see, {\it e.g.}, \cite{Contino:2016jqw}).
For example, in holographic composite Higgs models~\cite{Agashe:2004rs}, the mass scale is set by the Kaluza-Klein (KK) 
mass $M \sim e^{-k r \pi} k \equiv M_{\rm KK}$ and the coupling-strength is given by the rescaled five-dimensional 
gauge coupling $g_\ast \sim g_5/\sqrt{2 \pi r}$, where $k$ is the AdS$_5$ curvature and $r$ the compactification 
length of the fifth dimension.

In basic examples of integrating out a narrow resonance with a universal coupling to the SM, 
the couplings entering the coefficients $c^{(D)}_i$ can in many cases indeed be identified with the 
single NP coupling $g_\ast$. On the other hand, in more 
complicated scenarios, interactions of the SM sector with NP might involve additional small (mixing) parameters 
and different operators might come with different effective couplings. In the classical SILH, in fact operators 
involving gauge bosons (or light fermions) feature in general smaller couplings $g_V < g_\ast$, 
due to the non-maximal mixing of the corresponding fields with the strong sector.\footnote{
Note also that, due to the assumption of minimal coupling as well as symmetry considerations, some operators 
in the SILH feature further (loop) suppression factors \cite{Giudice:2007fh,Liu:2016idz}.
}
Such suppressions can however be lifted in scenarios that complement the SILH extension of the SM. In a 
setup of vector-compositeness, dubbed {\it Remedios} \cite{Liu:2016idz}, also gauge bosons couple in certain 
cases with the same strength to the strong sector as the Higgs, $g_V = g_\ast$.  
We will denote variants of well-known scenarios that feature just the latter
difference with a bar, {\it i.e.}, $\overline{\rm SILH}$, in the case discussed before.  Beyond that, in a general description 
of a light Higgs, without identifying it with a Goldstone boson as in the SILH, but rather assuming the smallness 
of the electroweak scale is due to some other mechanism or an accident -- the ALH -- (loop) suppression factors 
of the SILH are not present (see \cite{Liu:2016idz}). The concrete scaling of operators in these two basic frameworks of BSM 
physics (including their 'Remedios' versions\footnote{We assume the Remedios+MCHM of \cite{Liu:2016idz},
however the MCHM-like scaling entering in the last five columns of Table~\ref{tab:cor} is not crucial for the following analysis.}) is summarized 
in Table~\ref{tab:cor}, together with the example of integrating 
out a 
scalar $SU(2)_L$ doublet $S(1,2)_{1/2}$ with hypercharge $Y=1/2$ 
\cite{deBlas:2014mba,Henning:2014wua}.\footnote{We thus 
consider 
$\lambda_\varphi = g_\ast^2$ and $y_\varphi^f = y_f$.} The corresponding 
operators are defined in Table \ref{tab:ops}, where we employ the SILH basis (with an adapted 
normalization)~\cite{Giudice:2007fh,Liu:2016idz}.
Note that we always assume minimal flavor violation (MFV) to be at work, 
dictating the flavor structure of the operators \cite{DAmbrosio:2002vsn}, such as the coefficients 
$\lambda^{4\!f}$ or the yukawa couplings entering Table~\ref{tab:cor}. Four-fermion operators under 
investigation below will be assumed to feature left-handed quark currents
for concreteness, resulting in $\lambda^{4\!f}\to V_{tb} V_{ts}^\ast$ 
in the case of $bs$ transitions to leading approximation.\footnote{This is for illustration and the 
generalization to other operators that are generated in the scenarios is straightforward. Note that
MFV dictates the {\it ratios} of couplings accompanying different quark currents and
we neglect subleading terms, suppressed by powers of $y_i^2/y_t^2 \ll 1$. Moreover, it is assumed that chiral-symmetry breaking
effects are mediated by SM-like Yukawa couplings, leading to the appearance of $y_f$ in $c_{y_f}$ (or that the same physics that generates the $D\!=\!4$ 
Yukawa couplings also generates ${\cal O}_{y_f}$, which is true in many BSM scenarios,
such as in composite Higgs models). We will also comment on the concept of {\it partial compositenss}
further below.} Finally, we will comment on the scenario where $g_V \to g_{\rm SM}$ in footnote \ref{fn:SILH}.

\definecolor{gray2}{rgb}{0.95,0.95,0.95}
\definecolor{gray3}{rgb}{0.92,0.92,0.92}
\definecolor{gray4}{rgb}{0.75,0.75,0.75}
\newcolumntype{g}{>{\columncolor{gray2}}c}
\begin{table}[t]
\centering
\begin{tabular}{|c||c|c|c|c||c|c|c|c|c|}
\hline
  & ${\cal O}_{y_f}$ &  ${\cal O}_{4f}$ & ${\cal O}_6$ &${\cal O}_{3W,3G}$ & ${\cal O}_{BB,GG}$ &\!\!${\cal O}_{W,B}$\!\!&\!\!${\cal O}_{2W,2B,2G}$\!\!&\!\!${\cal O}_{HW,HB}$\!\!&\!${\cal O}_H$\!\\
\hline
\hline
$ \overline{\rm SILH} $ & $ \cellcolor{gray3}\ \marktopleft{a1} \! y_f g_\ast^2 $ & \cellcolor{gray3}  $\! \text{\footnotesize $\lambda^{4\!f}$} g_\ast^2\ \!$ &  \cellcolor{gray2} $\!\frac{y_t^2}{16 \pi^2} g_\ast^4$ &  \cellcolor{gray2}  $\! \frac{g_\ast^2}{16 \pi^2} g_\ast$
&  $\frac{y_t^2}{16 \pi^2} g_V^2 \!$ & $g_V$ & $1$  & $\!\! \frac{g_\ast^2}{16 \pi^2} (g,\!g^\prime)\!\!$  & $g_\ast^2$  \\
\hline
SILH   & $ \cellcolor{gray3}  y_f g_\ast^2$  & \cellcolor{gray3} $\! \text{\footnotesize $\lambda^{4\!f}$} g_\ast^2\ \!$  & \cellcolor{gray2} $\! \frac{y_t^2}{16 \pi^2} g_\ast^4$ \markbottomright{a1} & $\, \frac{g_V^2}{16 \pi^2} g_V$   &  $\frac{y_t^2}{16 \pi^2} g_V^2\!$ & $g_V$ & $\frac{g_V^2}{g_\ast^2}$ & $\frac{g_\ast^2}{16 \pi^2} g_V$ &  $g_\ast^2$  \\
\hline
$\, \overline{\rm ALH}\, $ & $ \cellcolor{gray3} \ \marktopleft{b2} \! y_f g_\ast^2$ &\cellcolor{gray3} $\! \text{\footnotesize $\lambda^{4\!f}$} g_\ast^2\ \!$ & \cellcolor{gray3}  $g_\ast^4$ & \cellcolor{gray3}  $g_\ast$ &  $g_V^2$ &  $g_V$ & $1$  & $g,g^\prime$  & $g_\ast^2$  \\
\hline
ALH & $ \cellcolor{gray3} y_f g_\ast^2$ &  \cellcolor{gray3}  $\! \text{\footnotesize $\lambda^{4\!f}$} g_\ast^2\ \!$ & \cellcolor{gray3} $g_\ast^4$ & $\frac{g_V^2}{g_\ast^2} g_V$ &   $g_V^2$ & $g_V$ & $\frac{g_V^2}{g_\ast^2}$  & $g_V$ & $g_\ast^2$  \\
\hline
\end{tabular}
\raggedright
\begin{tabular}{|c||c|c|c|c||}
$ \hspace{0.8mm} \int \! {[\phi]}$  \hspace{0.6mm}  & \hspace{-1.mm} $ \cellcolor{gray3} y_f g_\ast^2  $ 
&  \cellcolor{gray3} $\! \text{\footnotesize $\lambda^{4\!f}$} g_\ast^2 \ \! $   & \cellcolor{gray3} $\hspace{3.6mm} 
g_\ast^4 \hspace{0.1mm} $  
\hspace{-20mm} \markbottomrightblue{b2} \hspace{20mm} & $ \frac{g^2}{16 \pi^2}\frac{g}{60} \ $  \\
\hline
\end{tabular} \qquad \qquad \qquad $\lesssim (4 \pi)^{-2}$
\caption{\label{tab:cor} Scaling of the coefficients of the various $D=6$ operators in terms of couplings, in the framework of the $\overline{\rm SILH}$ (first line), 
the ordinary SILH (second line), the $\overline{\rm ALH}$ (third line), the ALH (fourth line), and when integrating out a 
narrow 
scalar $S(1,2)_{1/2}$ (fifth line). $\lambda^{4\!f}$ denotes
the flavor structure, see text for details.
}
\end{table}

\begin{table}[t]
\centering
\begin{tabular}{|c|}
\hline
  ${\cal O}_{y_f} = |H|^2 \bar f_L H f_R$ \\[0.5mm]
\hline
${\cal O}_{4f} = \bar f \gamma^\mu f\, \bar f \gamma_\mu f $ \\[0.5mm]
\hline
${\cal O}_6 = |H|^6$ \\[0.5mm]
\hline
${\cal O}_{3V} = \frac 1 {3 !} F_{abc} V_\mu^{a\nu}V_{\nu\rho}^b V^{c \rho \mu}$ \\[0.5mm]
\hline
\end{tabular}
\begin{tabular}{|c|}
\hline
${\cal O}_{VV} = |H|^2 V_{\mu \nu}^a V^{a \mu \nu}$ \\[0.5mm]
\hline
${\cal O}_V = \frac i 2 (H^\dagger \sigma^a \overleftrightarrow{D^\mu} H) D^\nu W_{\mu\nu}^a  $ \\[0.5mm]
\hline
${\cal O}_{2V} = -  \frac 1 2(D_\rho V_{\mu\nu}^a)^2 $ \\[0.5mm]
\hline
${\cal O}_{HV} = i(D^\mu H)^\dagger \sigma^a (D^\nu H) V_{\mu\nu}^a $ \\[0.5mm]
\hline
${\cal O}_H = \frac 1 2(\partial^\mu |H|^2)^2 $ \\[0.5mm]
\hline
\end{tabular}
\caption{\label{tab:ops} The operators under consideration, where {\small \mbox{$3V\!=\!3W,3G;$} \mbox{$VV\!=\!BB,GG;$} 
\mbox{$V\!=\!B,W;$} \mbox{$2V\!=\!2B,2W,2G;$} \mbox{$HV\!=\!HB,HW$}}. Moreover, $a,b,c=1,..,8;\ 1,..,3;\ \varnothing$, 
for $SU(3),SU(2)_L,U(1)_Y$, respectively, with $F_{abc}=f_{abc},\epsilon_{abc},1$ the corresponding 
structure constants (and clearly the $\sigma^a$ matrices absent in the case of $V\!=\!B$ as well as 
$D=\partial$ when acting on $B_{\mu\nu}$). Note that $f$ denotes, schematically, fermion fields.}
\end{table}

We note from Table \ref{tab:cor} that in general strong correlations (via $g_\ast$) exist between the operators 
${\cal O}_{y_f}$, ${\cal O}_{4f}$, ${\cal O}_6$ (and ${\cal O}_{3V}$ in the case of the Remedios setup), which 
have the same form in many scenarios, as emphasized by the shades of gray. We will thus consider measurable quantities 
that are transparently related to these operators in the following. Two diverse scenarios (visualized by orange and blue 
dashed lines) can be identified, which can be mapped to two benchmarks for the analysis of this article, capturing basically 
all models at hand concerning the class of operators under consideration.
They are distinguished by the assumption whether ${\cal O}_6$ 
is tree or loop generated and summarized in Table~\ref{tab:cor2}, denoted by capital letters ${\bf A}$ and ${\bf B}$.\footnote{\label{fn:SILH}The given scaling in the case of ${\cal O}_{3V}$ holds only in the Remedios setups, where $g_V=g_\ast$. While other 
correlations are rather robust, those including this operator will depend on this assumption. 
Note moreover, that in the original version of the SILH 
proposal~\cite{Giudice:2007fh} vectors were assumed to be weakly coupled - featuring exactly the SM gauge couplings - which would lead to the replacement $g_V \to g,g^\prime,g_s$ in columns $4\!-\!8$ in the second line of Table \ref{tab:cor},
for $V=W,B,G$, respectively. Clearly, making such a specific assumption on the value of couplings entering the NP 
terms could let us hope to be able to determine $M$, in this particular scenario, via processes involving gauge bosons. 
The analysis presented here on the one hand considers the generalized context where all NP terms could
appear with a NP coupling - $g_V$ in the case of gauge bosons - (focusing on operators with~a
'universal' scaling), but beyond that particularly envisages effects in operators 
without gauge bosons, 
${\cal O}_{y_f},{\cal O}_{4f}$, and ${\cal O}_6$, with still ample room for NP. 
Moreover, TGC measurements and Higgs decays are often only sensitive to combinations of 
operators with different scaling, like $c_{W,B}$ and $c_{2W,2B}$.}

In the following section, we will detail how these scenarios can be tested by studying correlations between observables.
In particular, we will discuss which patters of deviations are expected and how they would allow us to access directly NP
masses, beyond collider reach, and which patterns would exclude given UV paradigms.
In fact, as we will work out below, accessing two operators at a time will allow for a 'model-independent' direct 
determination of the NP mass $M$ (without making an assumption
for the coupling $g_\ast$), valid in a large class of NP frameworks - with the only remaining freedom 
being the question whether one is in Scenario~{\bf A} or in Scenario~{\bf B}.

The latter information can however be obtained by including a third observable, 
a procedure that
we will explicitly go through at the end of this article.
Indeed, considering simultaneously measurements of the 
coefficients of two operators will determine both $M$ and $g_\ast$, which leads to a distinct
prediction for the remaining coefficients in both scenarios, which can be confronted with 
bounds, eventually allowing us to determine which of the scenarios is viable, and which is 
excluded. 


\begin{table}[t]
\centering
\begin{tabular}{|c||c|c|c|c|}
\hline
  & ${\cal O}_{y_f}$ &  ${\cal O}_{4f}$ & ${\cal O}_6$ &${\cal O}_{3W,3G}$ \\
\hline
\hline
$ {\bf A} $
& $ y_f g_\ast^2$ & $\lambda^{4\!f} g_\ast^2$ & $g_\ast^4$ & $g_\ast$ \\
\hline
$ {\bf B}$  & $ y_f g_\ast^2$ & $\lambda^{4\!f} g_\ast^2$ &  $\frac{y_t^2}{16 \pi^2} g_\ast^4$ & $\frac{g_\ast^2}{16 \pi^2} g_\ast$\\
\hline
\end{tabular}
\caption{\label{tab:cor2} Operators that can be basically divided in two classes of scalings, {\it i.e.}, ${\bf A}$:~ALH-like
and ${\bf B}$:~SILH-like. See text for details.}
\end{table}

\section{Analysis + Discussion}
\label{sec:ana}

We will now study in detail how measurable quantities, that depend in a simple way on the operators identified in 
Section \ref{sec:setup}, can be employed to unveil the mass of new states, even in case the available energy does not
suffice to produce them directly. Let us start by exploring how simultaneous 
measurements of more than one such quantity can be used to lift the flat direction in $g_\ast$ vs. $M$ and 
examine the resulting sensitivity on $M$. 

Consider two operators ${\cal O}_1$ and ${\cal O}_2$, whose coefficients, $c_1 \sim g_\ast^{y_1}/M^2$ and 
$c_2 \sim g_\ast^{y_2}/M^2$, feature a different scaling in $g_\ast$, $y_1 \neq y_2$. Now, assume these 
coefficients are extracted from measurements, resulting in $c_1 = X_1$ and $c_2 = X_2$,
with $X_{1,2}$ featuring mass dimensions $D=-2$. We can now solve the simple system of equations
(neglecting loop factors, which can be implemented trivially)
\begin{equation}
\label{eq:sys}
\frac{g_\ast^{y_1}}{M^2} \sim X_1 \,,\quad \frac{g_\ast^{y_2}}{M^2} \sim X_2
\end{equation}
for $M$, leading (for $y_1 \neq y_2$) to
\begin{equation}
\label{eq:rel}
M \sim \left(\frac{X_1^{y_2}}{X_2^{y_1}}\right)^{\frac{1}{2(y_1-y_2)}}\,.
\end{equation}
We observe that in general the best sensitivity regarding NP masses results from studying pairs of operators 
that feature powers of $g_\ast$ that are large
for each of the operators, but close to each other, $\Delta_y\equiv |y_1-y_2| \sim 1$. 
In Scenario ${\bf B}$, this would correspond for example to the pair of operators ${\cal O}_6$ and ${\cal O}_{3V}$
(which are however loop suppressed, limiting the effects due to perturbativity), that feature $\Delta_y = 1$ and would lead to a sensitivity $M \approx (X_1^{3/2}/X_2^2)$. 
Moreover, the sensitivity increases if the operator that features the smaller power of $g_\ast$ becomes stronger constrained.
On the other hand, confronting measurements of $c_{y_f}$ and  $c_{4f}$
leads to no sensitivity at all in both scenarios - concerning the operators in Table \ref{tab:cor2}, at least a 
measurement of 
$c_6$ or $c_{3V}$ needs to be involved.

\begin{figure}[!t]
	\begin{center}
	\includegraphics[height=2.75in]{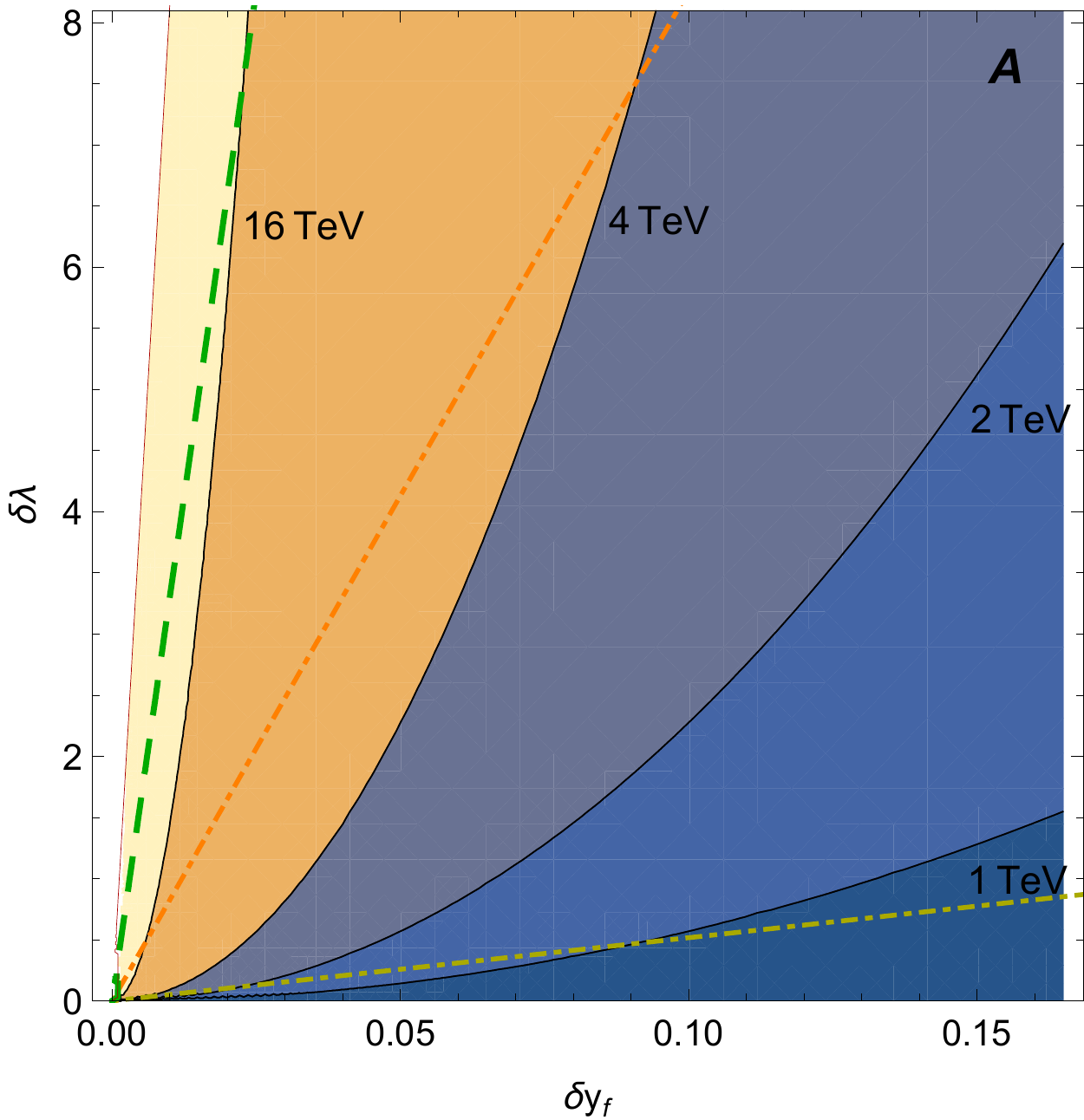}\qquad \includegraphics[height=2.75in]{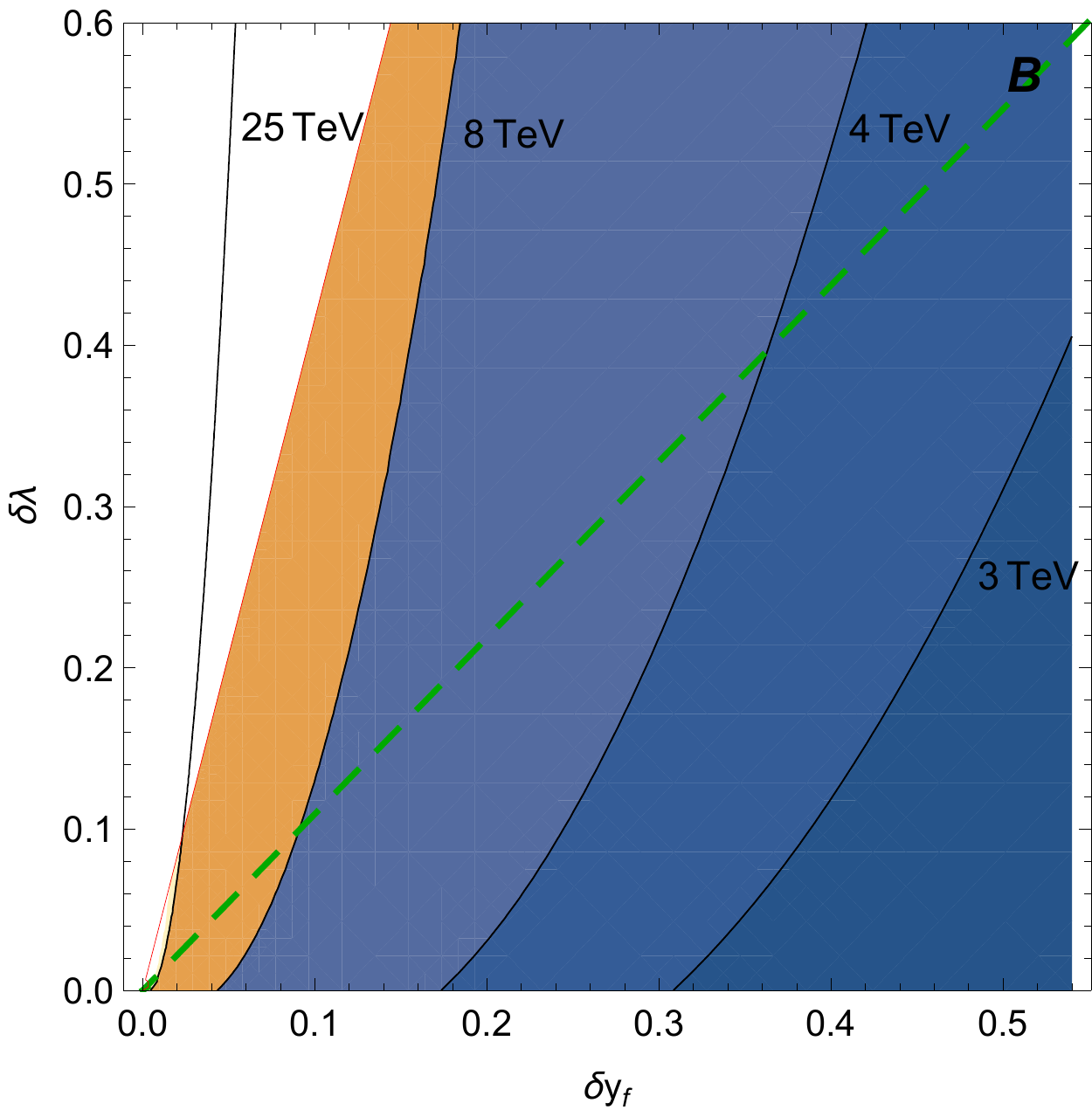}
	\caption{\label{fig:h3-Y} NP mass $M$ in dependence on the variation in the Yukawa couplings 
	($\delta y_f$) and in the triple-Higgs self coupling ($\delta \lambda$) in Scenario {\bf A} (left) and {\bf B} (right). 
	The colored lines denote NP couplings of $g_\ast = 1,4,8,4\pi$, respectively (from yellow to red). See text for details.
	}
	\end{center}
	\end{figure}

\begin{figure}[!t]
	\begin{center}
	\includegraphics[height=2.42in]{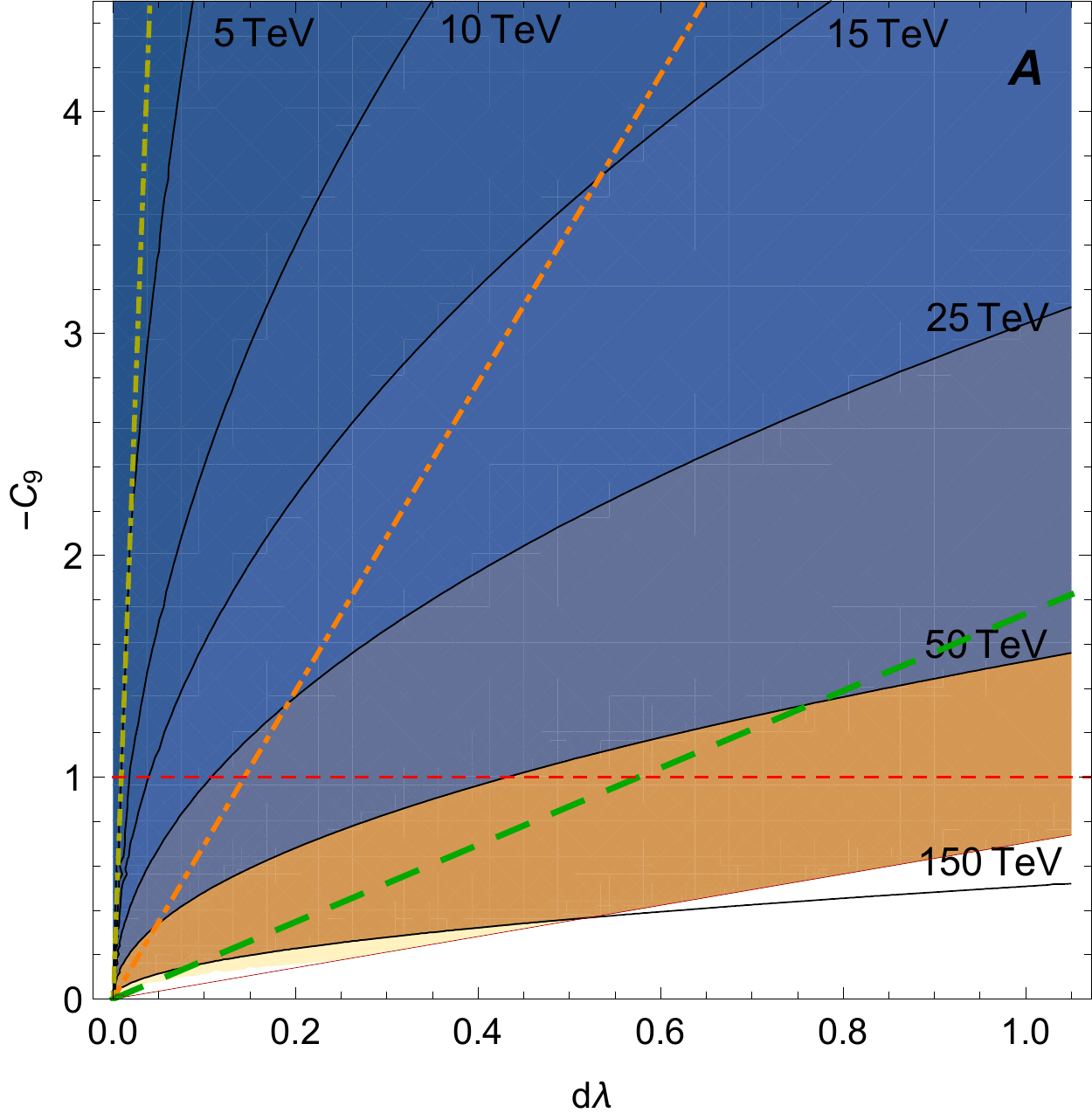}\qquad \includegraphics[height=2.42in]{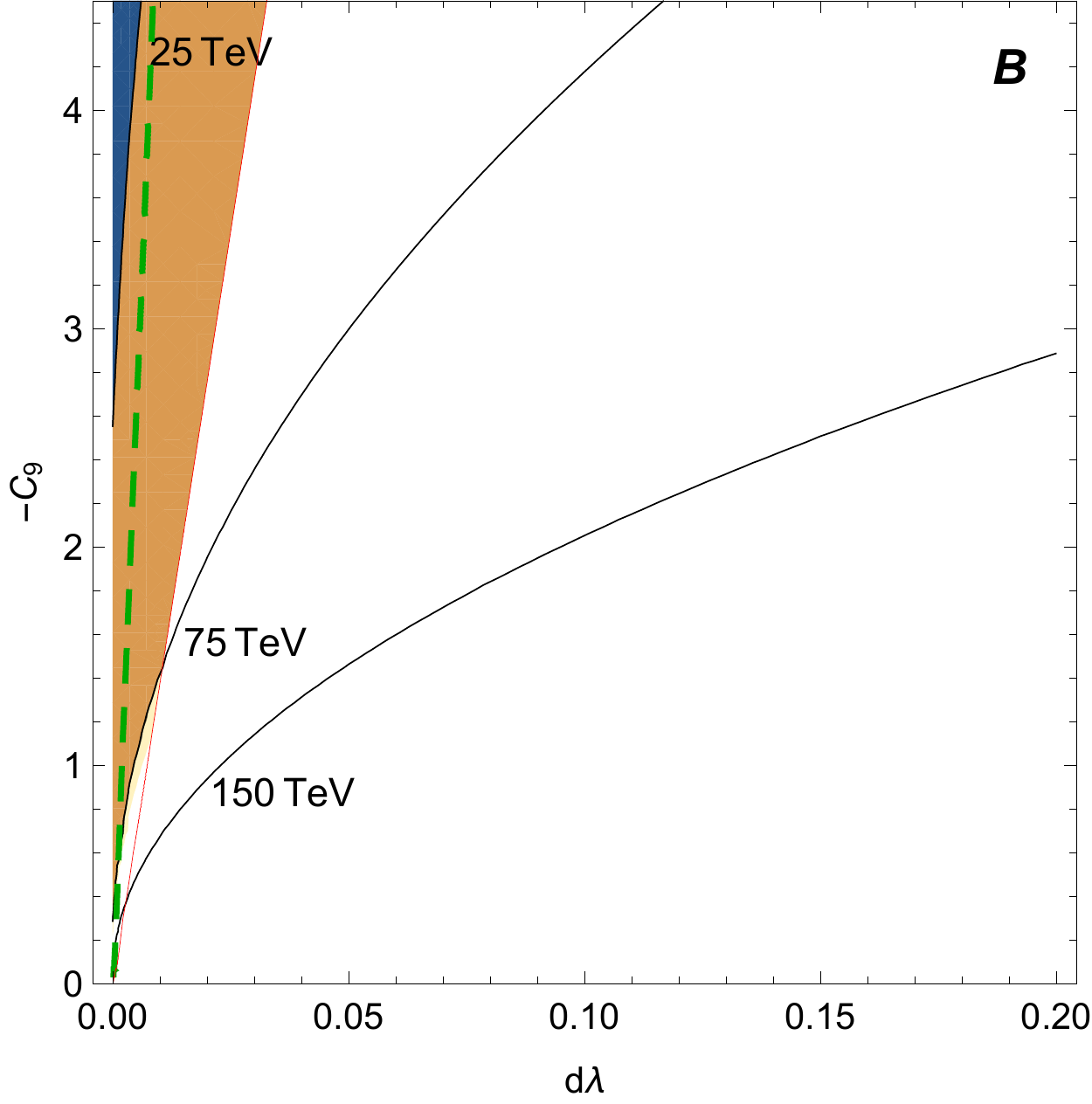}
	\\[3mm]
	\includegraphics[height=2.95in]{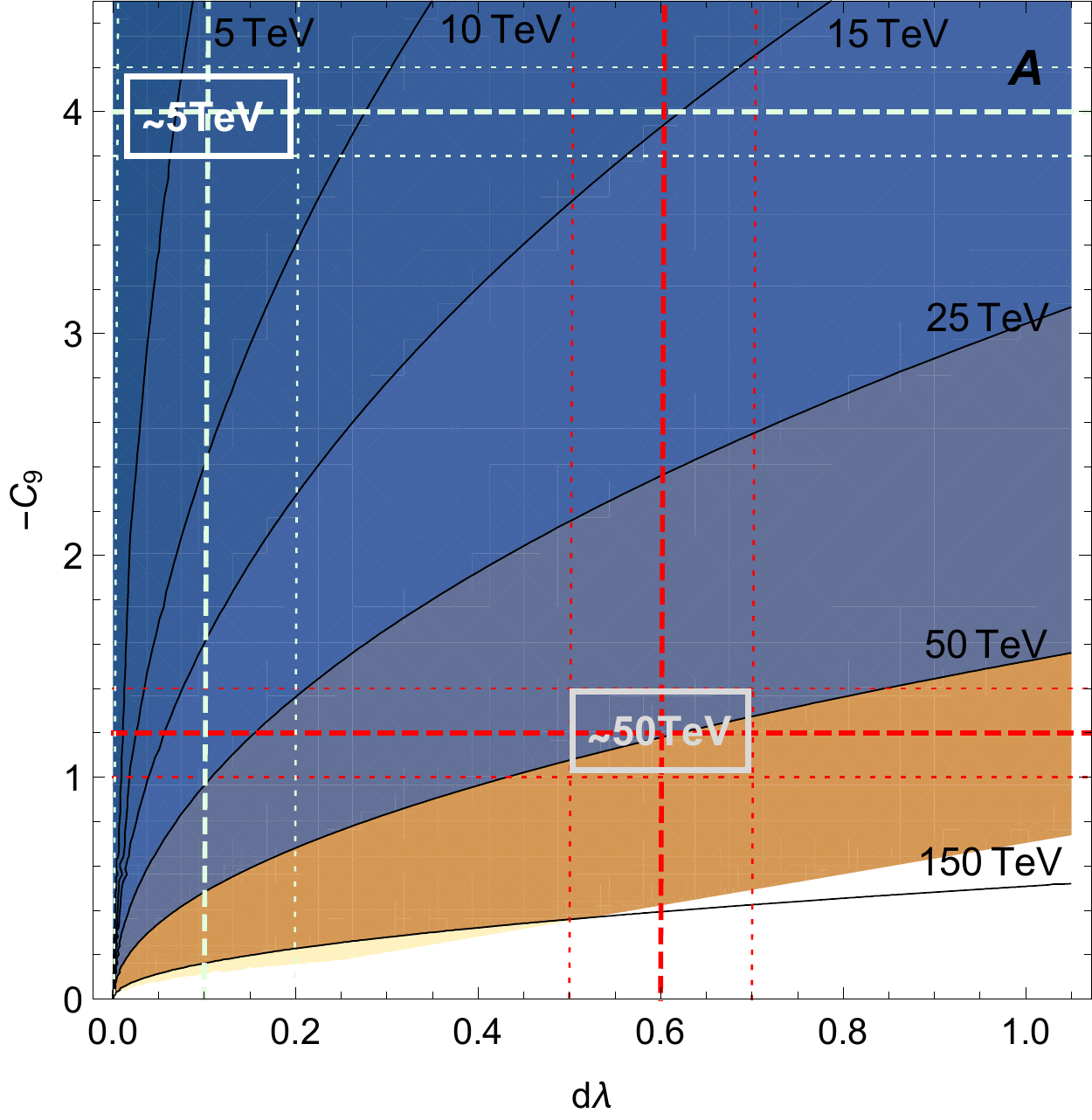}\qquad 
	\caption{\label{fig:c9-h3} NP mass $M$ in dependence on the variation in
	the triple-Higgs self coupling ($\delta \lambda$) and the coefficient of the four fermion
	operator ${\cal O}_9$ ($C_9$) in Scenario {\bf A} (upper panel, left) and {\bf B} (upper panel, right). 
	The colored lines denote NP couplings of $g_\ast = 1,4,8,4\pi$, respectively (from yellow to red). The lower
	panel includes two hypothetical measurements of a signal in the $\lambda_Z-c_9$ plane and illustrates the correspondingly
	extracted masses of new particles in Scenario {\bf A}. See text for details.}
	\end{center}
	\end{figure}

\begin{figure}[!t]
	\begin{center}
	\includegraphics[height=2.75in]{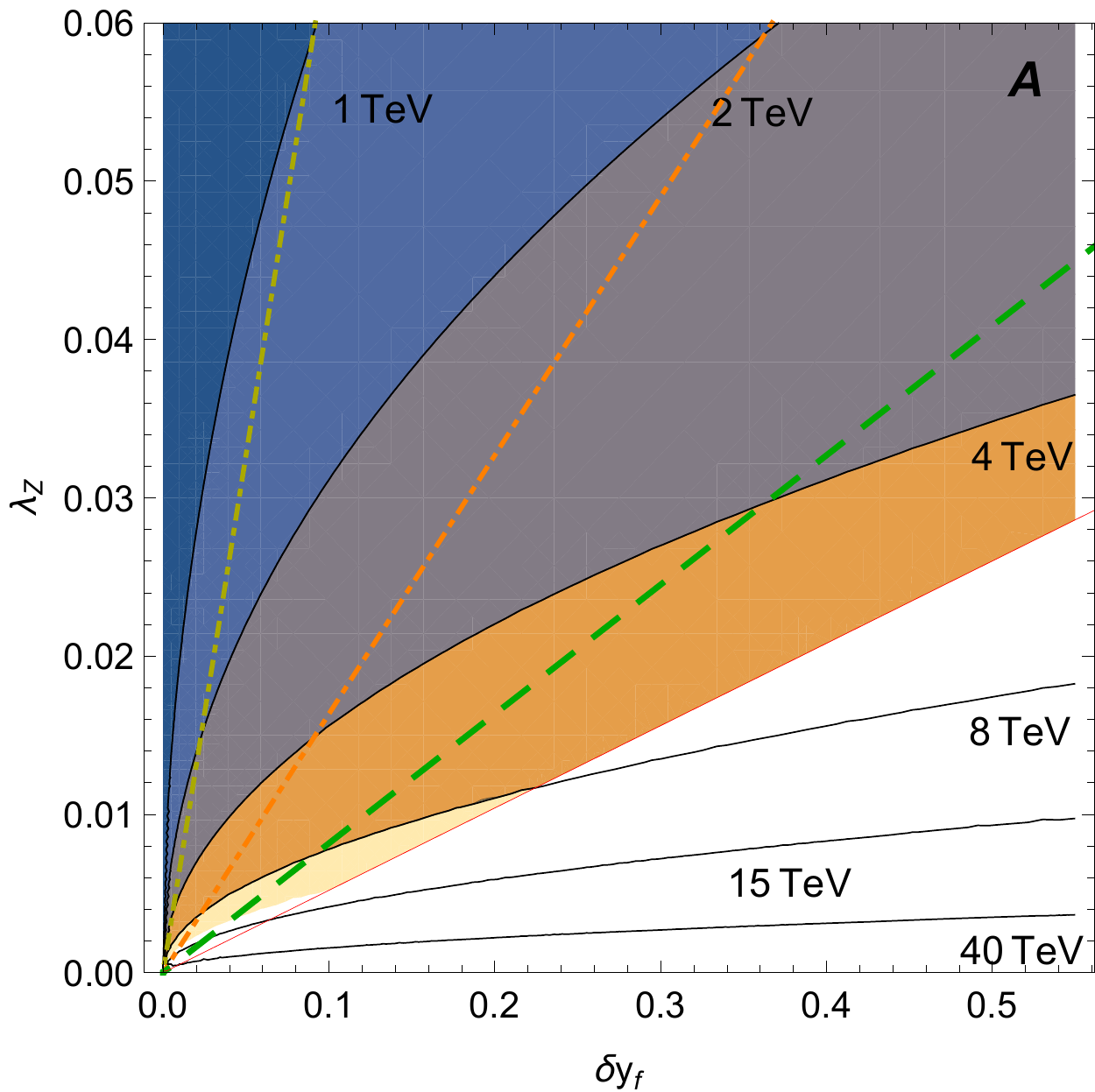}\:\qquad \includegraphics[height=2.75in]{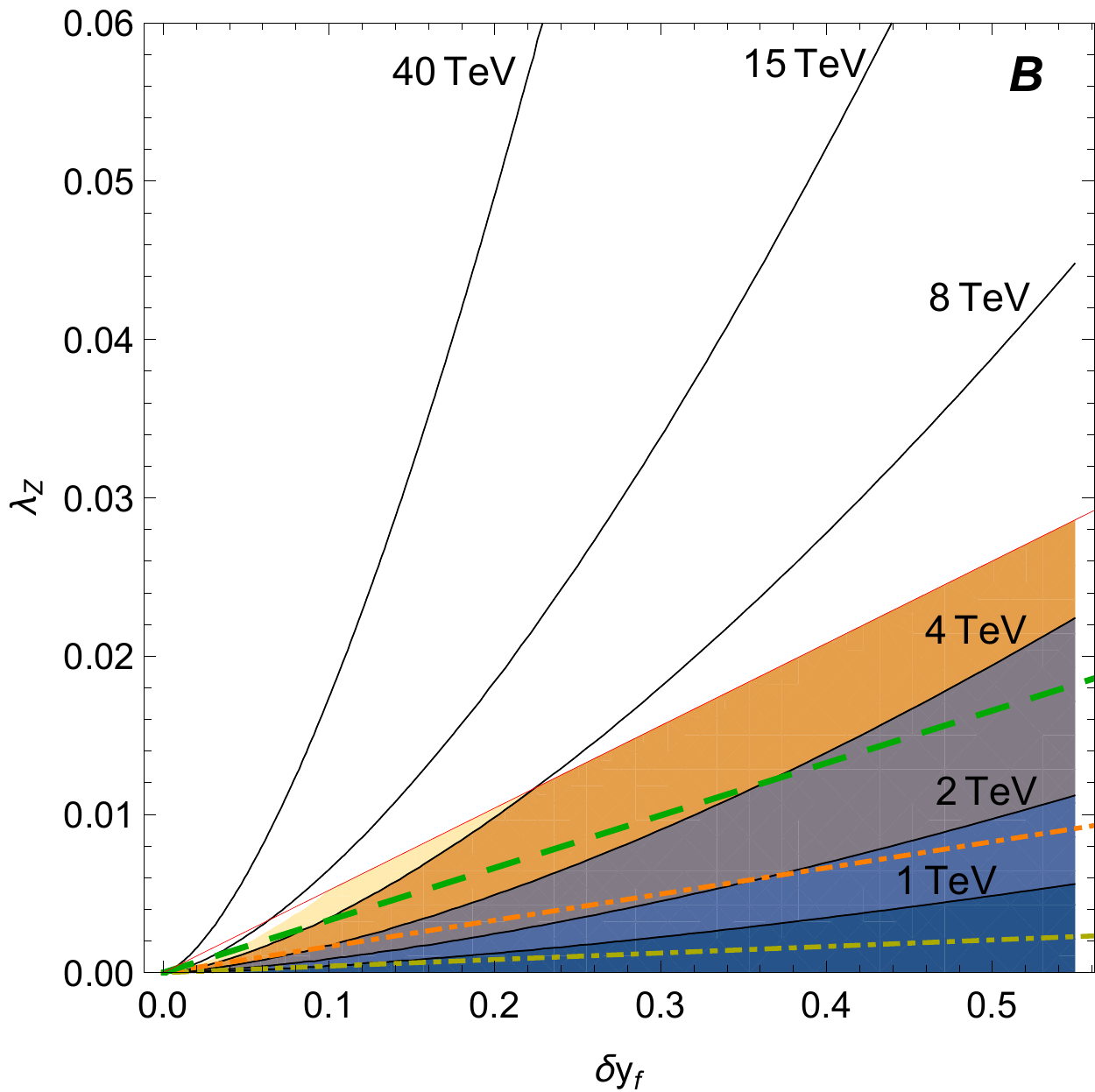}
	\\[3mm]
	\includegraphics[height=2.73in]{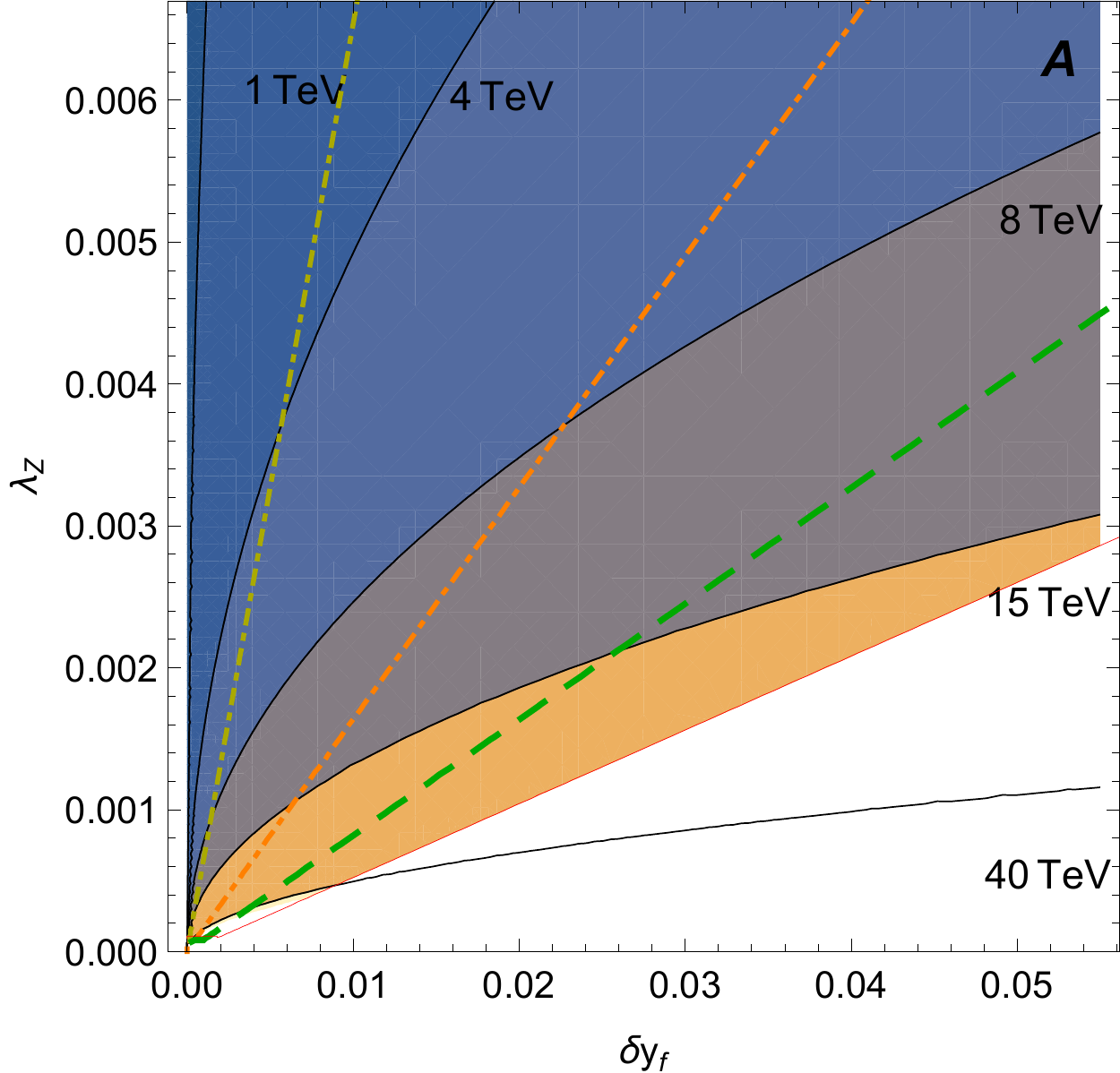}\,\quad \includegraphics[height=2.73in]{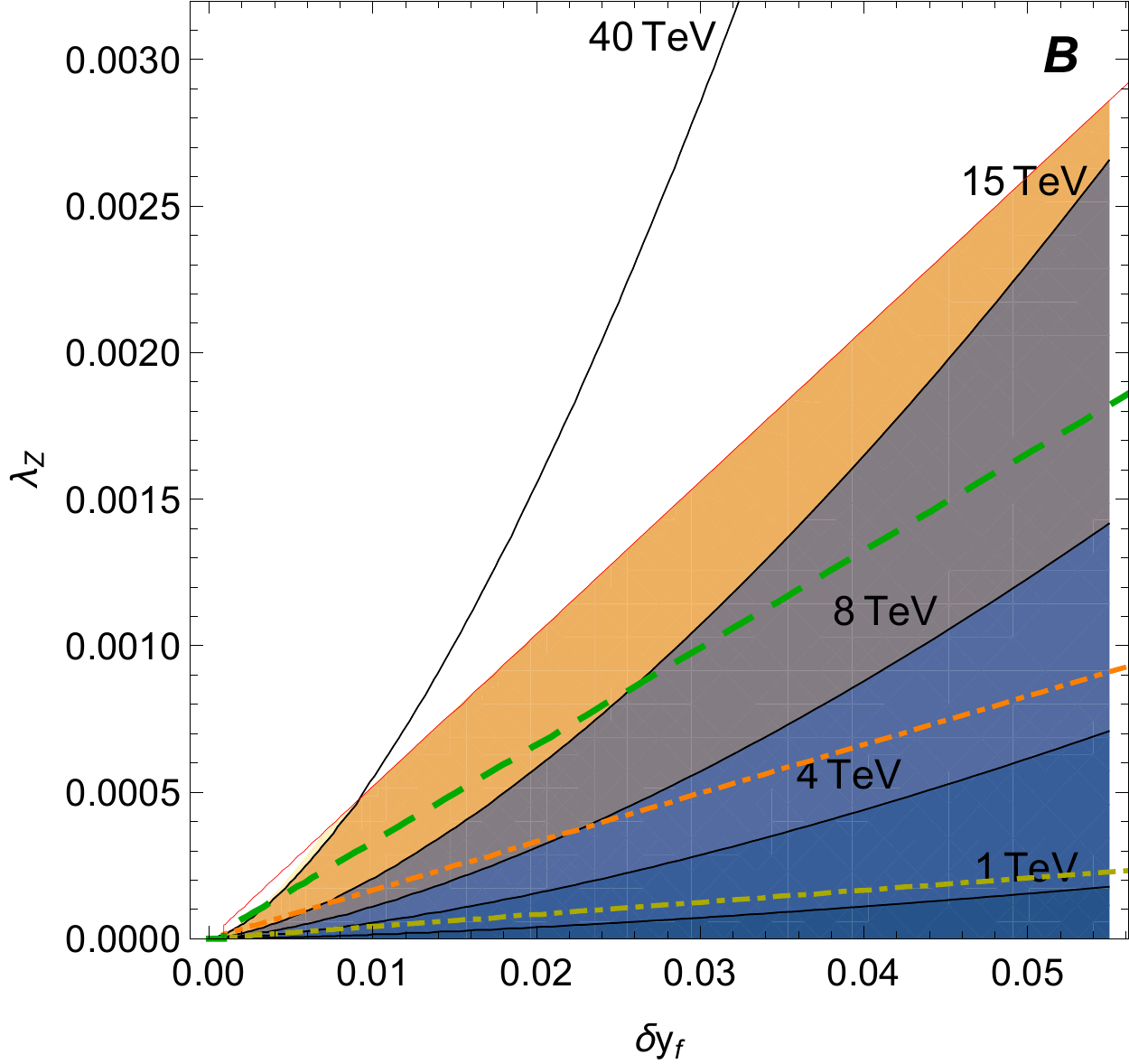}
	\caption{\label{fig:Y-TG} NP mass $M$ in dependence on the variation in the Yukawa couplings 
	($\delta y_f$) and the triple-gauge coupling ($\lambda_Z$) in Scenario {\bf A} (left) and {\bf B} (right). 
	The colored lines denote NP couplings of $g_\ast = 1,4,8,4\pi$, respectively (from yellow to red), and the lower
	panel shows a zoom into the plots of the upper panel. See text for details.}
	\end{center}
	\end{figure}

\begin{figure}[!t]
	\begin{center}
	\includegraphics[height=2.75in]{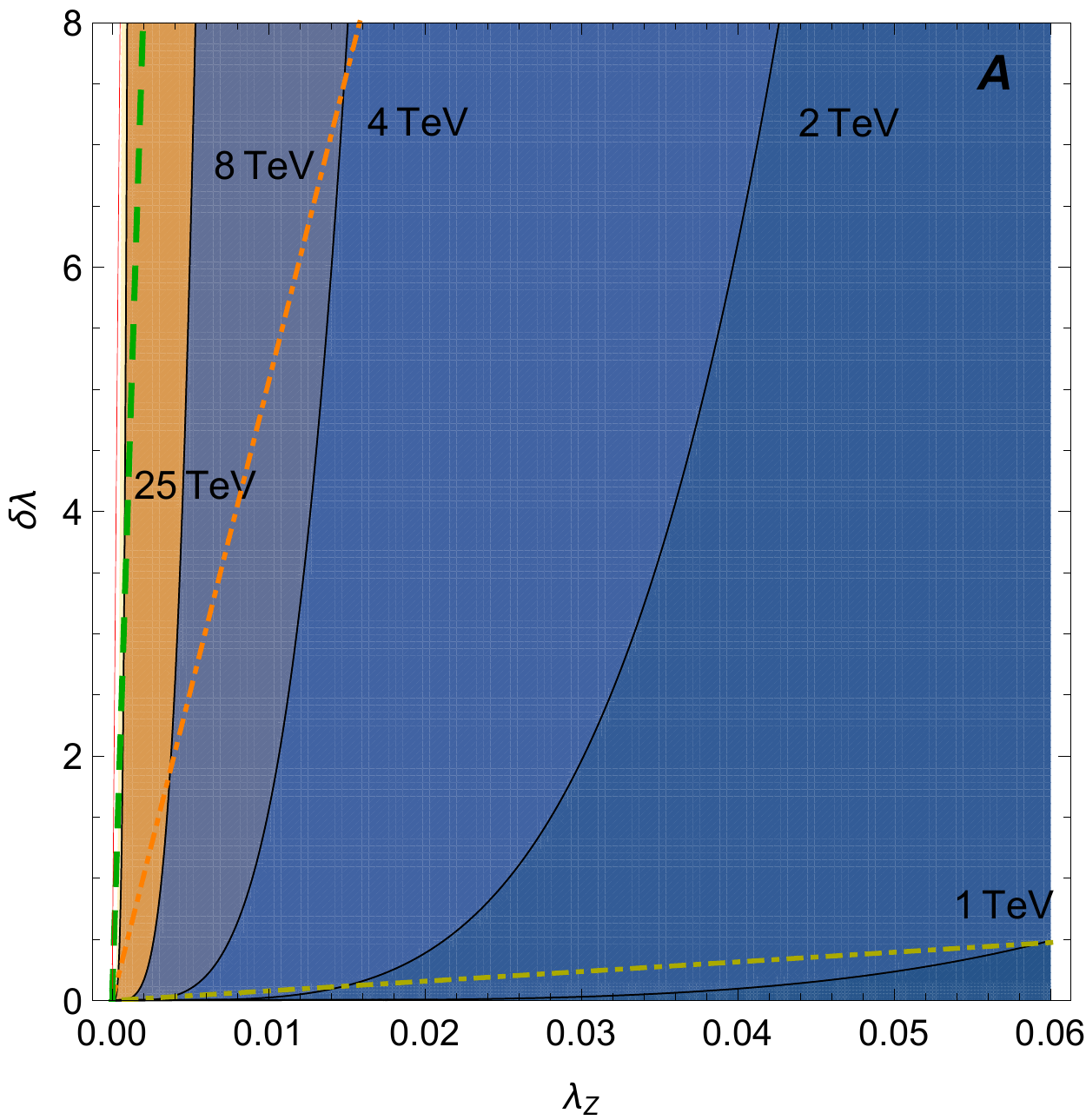}\qquad \includegraphics[height=2.75in]{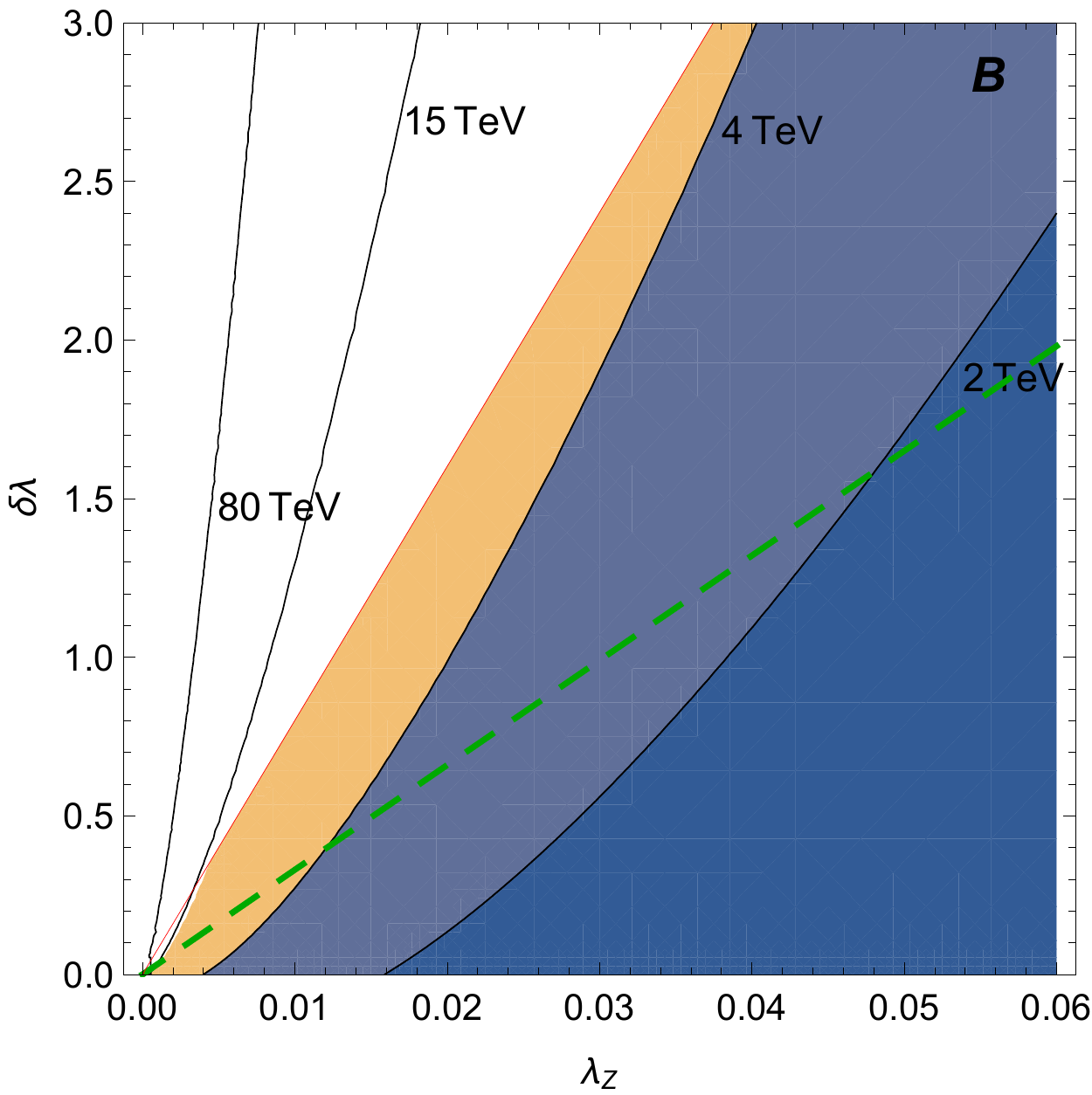}\\[2mm]
	\includegraphics[height=2.75in]{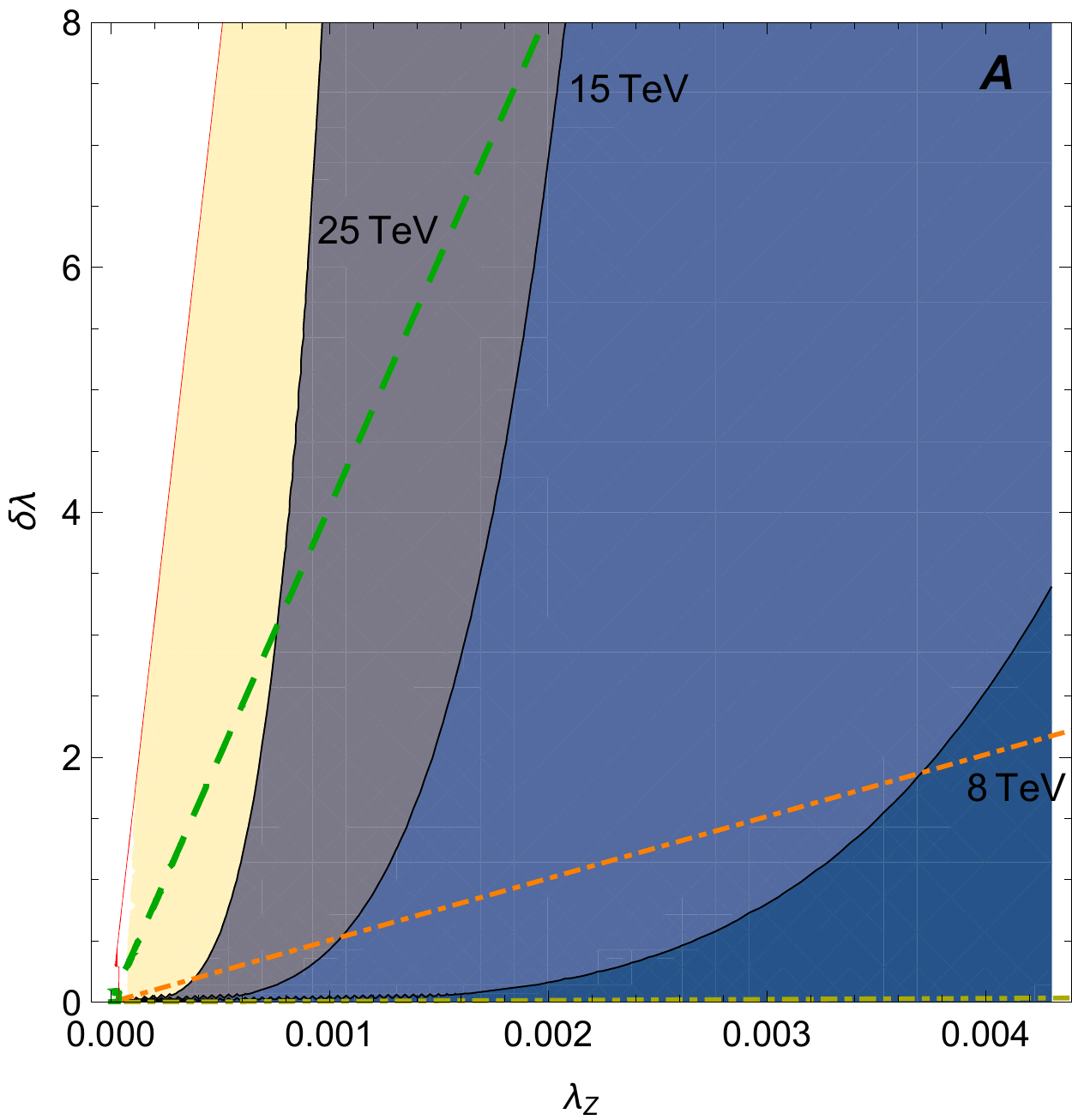}\qquad \includegraphics[height=2.75in]{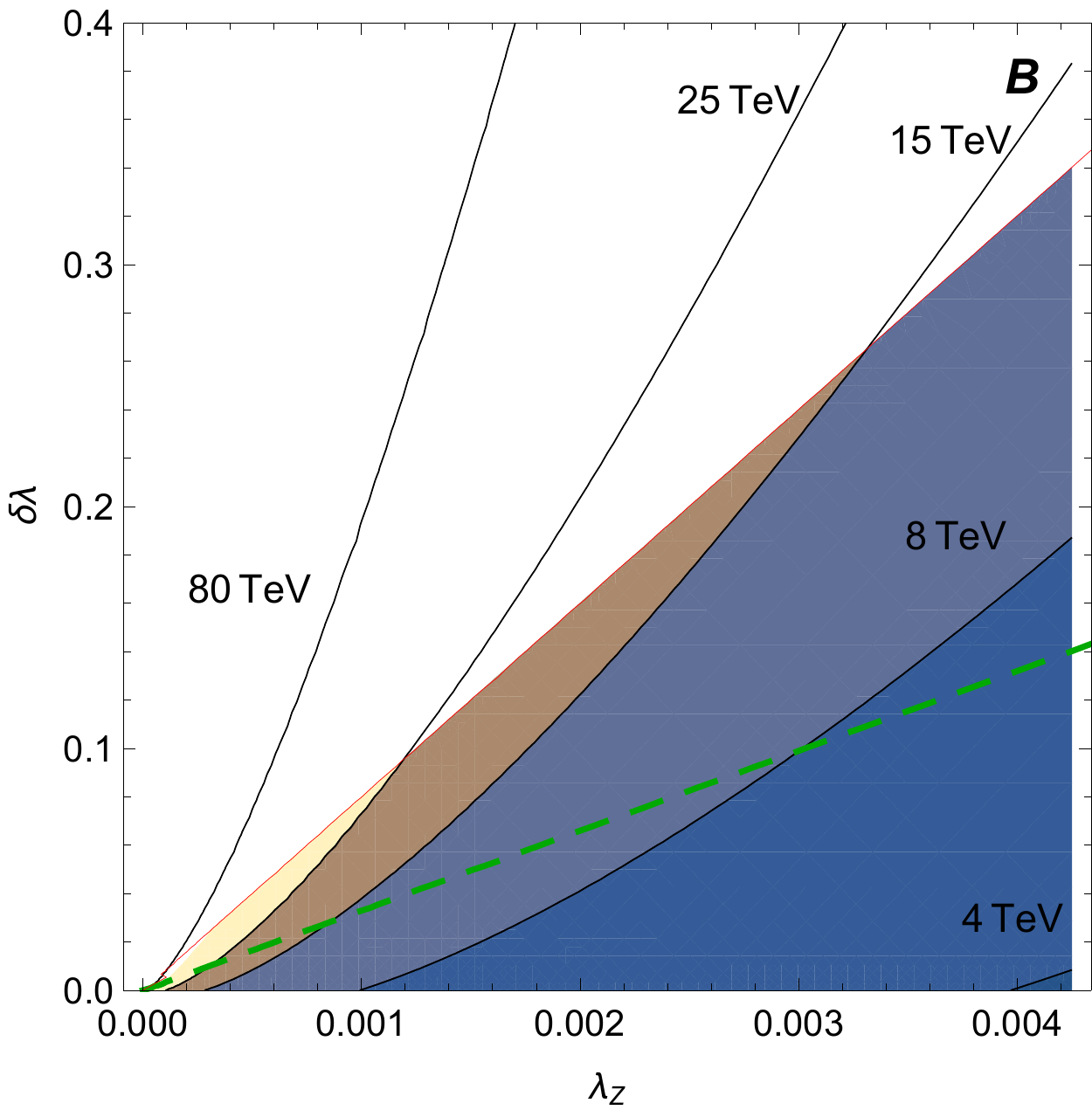}
	\caption{\label{fig:h3-TG} NP mass $M$ in dependence on the 
	the triple-gauge coupling ($\lambda_Z$) and the variation in the triple-Higgs self 
	coupling ($\delta \lambda$) in Scenario {\bf A} (left) and {\bf B} (right). 
	The colored lines denote NP couplings of $g_\ast = 1,4,8,4\pi$, respectively (from yellow to red), and the lower
	panel shows a zoom into the plots of the upper panel. See text for details.}
	\end{center}
	\end{figure}

\begin{figure}[!t]
	\begin{center}
	\includegraphics[height=2.75in]{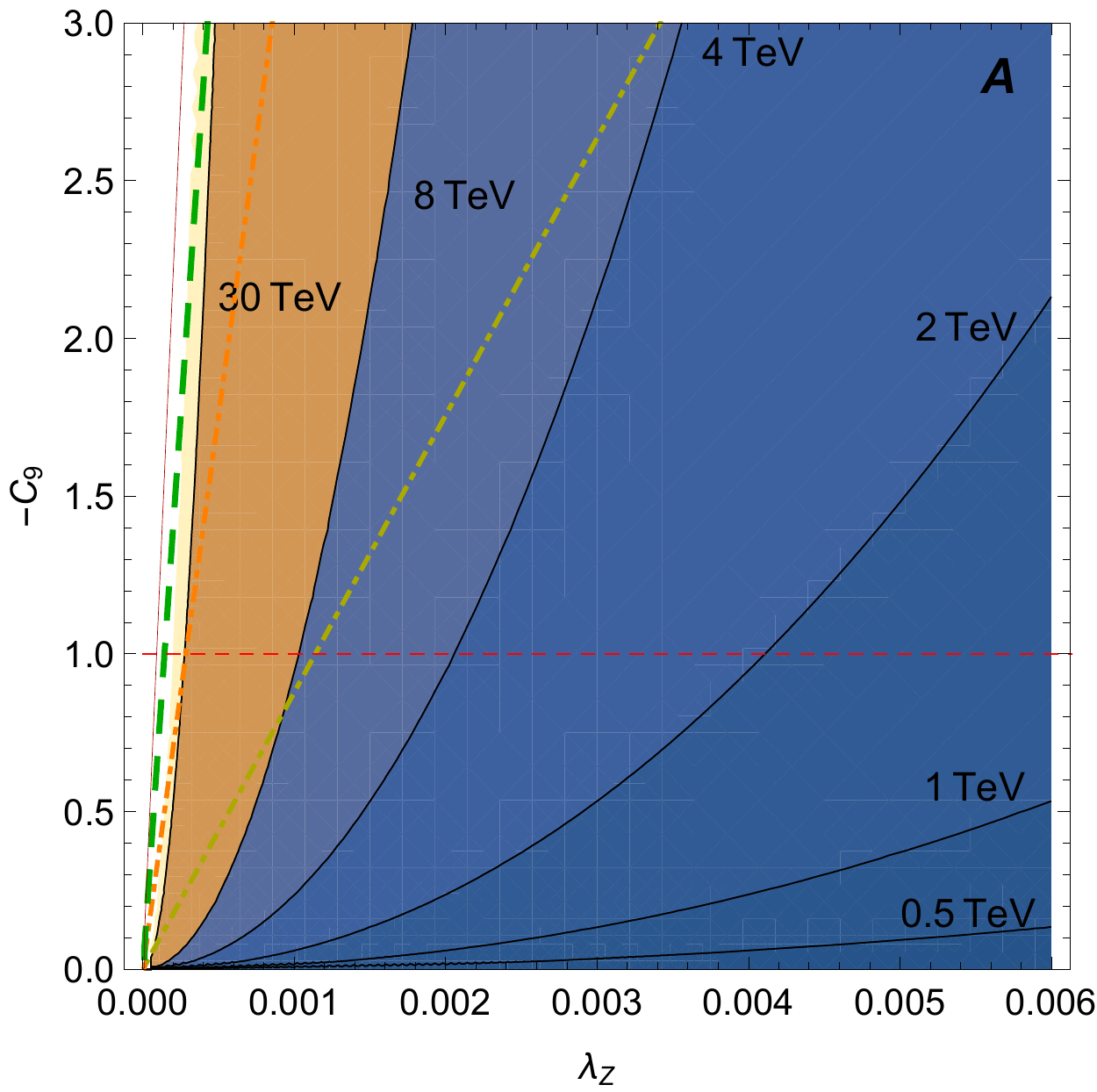}\qquad \includegraphics[height=2.75in]{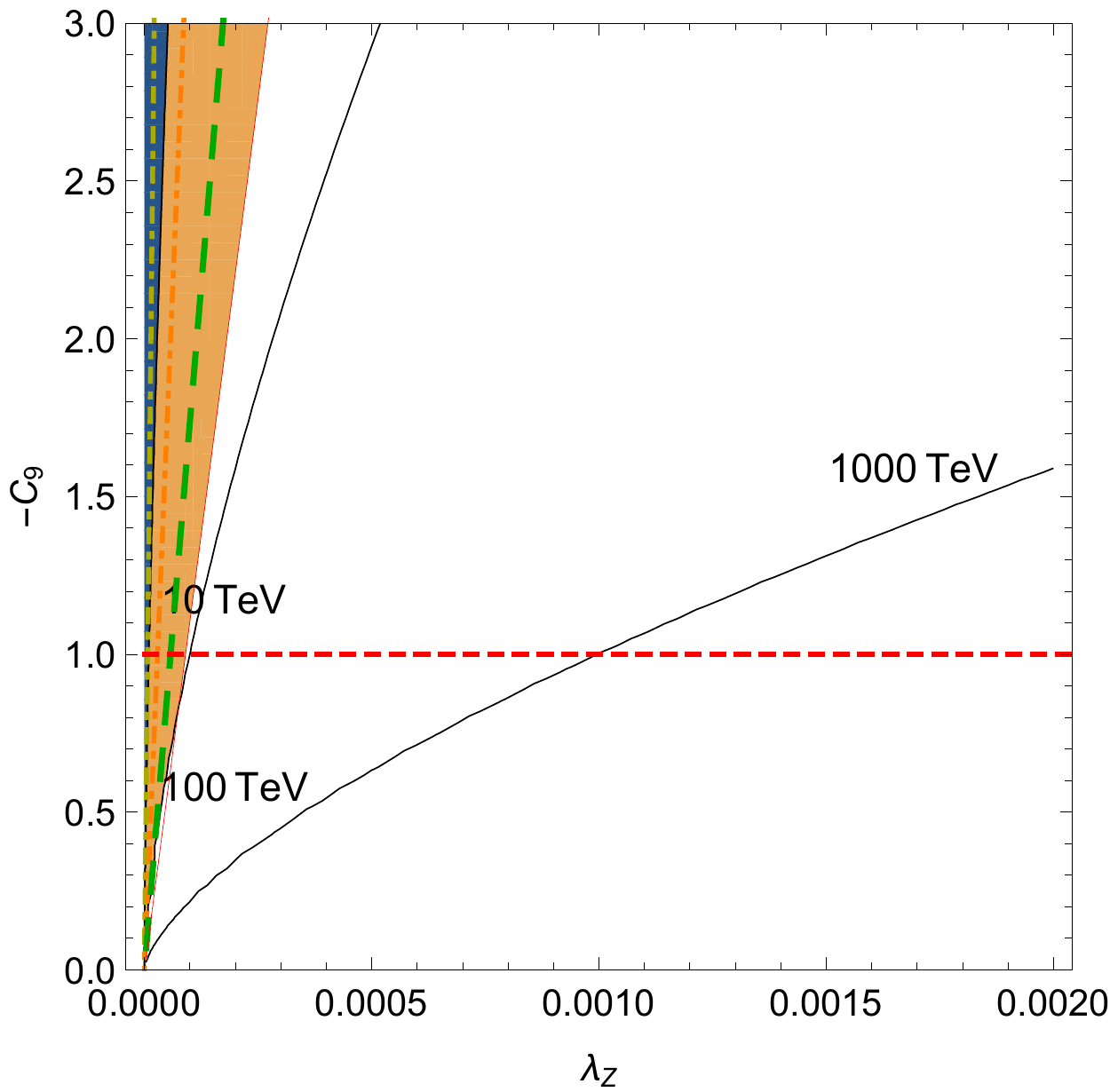}
	\caption{\label{fig:c9-TG} NP mass $M$ in dependence on 
	the triple-gauge coupling ($\lambda_Z$) and the coefficient of the four fermion operator ${\cal O}_9$ ($C_9$) 
	in Scenario {\bf A} (left) and {\bf B} (right). 
	The colored lines denote NP couplings of $g_\ast = 1,4,8,4\pi$, respectively (from yellow to red), and the horizontal
	dashed line depicts the currently preferred value of $C_9\sim -1$. See text for details.}
	\end{center}
	\end{figure}

\emph{A general procedure to determine the NP mass $M$, given a set of measurements, is now as follows}
\begin{enumerate}
\item{Consider a pair of measurements that determines two coefficients in Table \ref{tab:cor2} 
and solve for $M$ via relation (\ref{eq:rel}), employing the results presented in Figures~\ref{fig:h3-Y}-\ref{fig:c9-TG},
 both for Scenario {\bf A} and {\bf B}.}
\item{Solve for $g_\ast$ via eqs. (\ref{eq:sys}), derive predictions for the remaining pseudo-observables,
and confront them with measurements. Drop the scenario that is (experimentally)
excluded or inconsistent with EFT assumptions or perturbativity (too small $M$, too large $g_\ast$).}
\item{The remaining solution provides a direct estimate for the {\it mass} of new particles, completing
the SM.}
\end{enumerate}

While not involving $\lambda_Z$ at the beginning 
avoids assumptions regarding vector-boson couplings, it can help to discriminate between the different setups
in the end.
In fact, if at some point in the procedure above one encounters a significant contradiction, 
the corresponding underlying hypothesis ({\it e.g.}, the SILH with assumptions detailed before) 
can be excluded to be realized in nature.
	
We now move to a numerical study of the sensitivities to NP masses~\!$M$, considering (hypothetical) measurements
of 
\begin{enumerate}[1)]
	\item a relative shift in yukawa couplings $\delta {y_f}$
	\item the coefficient $C_9$ of the four-fermion operator\footnote{\label{fn:B} We take the operator ${\cal O}_9$ as an 
	example since several current anomalies hint to a non-zero value 
	of its coefficient, $C_9 \sim -1$ (see, {\it e.g.}, 	\cite{Descotes-Genon:2013wba,Gauld:2013qba,Altmannshofer:2017fio,DAmico:2017mtc}). Although not all 
	scenarios considered here can address the experimental observation in a fully consistent way,
	the case can serve as an illustrative example for the method in any case.} 
	${\cal O}_9 \equiv \frac{4 G_F}{\sqrt 2} V_{tb}V_{ts}^\ast \frac{\alpha}{4 \pi}  (\bar s_L \gamma_\mu b_L) 
	(\bar \ell \gamma^\mu \ell)$
	\item a relative deviation in the Higgs self coupling $\delta\lambda$
	\item the triple-gauge coupling (TGC) $\lambda_Z$.
\end{enumerate}
These quantities are related to the coefficients of the operators in Table~\ref{tab:cor2} as (see, {\it e.g.,} \cite{deFlorian:2016spz})\footnote{Note that $\delta y_f$ and $\delta \lambda$ also receive contributions from
a non-vanishing $c_H$. In the former case, these can be included simply by rescaling $\delta y_f$ by 
a factor of 3/2 \cite{deFlorian:2016spz}, which is accounted for in our numerical analysis (and for uniformity/simplicity
we adjust similarly $y_\varphi^f \to 3/2 y_f$ for the scalar resonance). 
In the latter case, for Scenario {\bf A} the effect is suppressed by the ratio of the (small)
SM-like trilinear self coupling over the NP coupling squared, $\lambda/g_\ast^2 $,
and thus basically negligible for $g_\ast \gtrsim 1$. Since the interesting parameter space just features
this range of couplings in all cases where $\delta \lambda$ is involved, this effect can be discarded.
In Scenario {\bf B}, we include the impact of $c_H$ by adding a term $- \frac 3 2 v^2 c_H$ to the third line of eq. (\ref{eq:rel2}).
}
\begin{equation}
\label{eq:rel2}
\begin{split}
\delta {y_f} & = v^3/(\sqrt 2 m_f)\  c_{y_f} \\
C_9 & = \frac{\sqrt 2 \pi}{\alpha\, G_F V_{tb}V_{ts}^\ast }\ c_{\text{\tiny $s_L\!b_L\!\ell \ell$}}\\
\delta\lambda & = 2 v^4/m_h^2\ c_6 \\
 \lambda_Z & = -6 g^2\ c_{3W}\,.
\end{split}
\end{equation}
In that context, recall that $c_{\text{\tiny $s_L\!b_L\!\ell \ell$}}$ is assumed to scale like 
$V_{tb} V_{ts}^\ast g_\ast^2$, respecting MFV (such a structure
is viable to allow for considerable effects in $C_9$, without being directly excluded from other measurements in flavor physics). 
A similar scaling holds in the case of partial compositeness, with the left-handed ($b$) quarks 
and leptons coupled significantly to the composite sector \cite{Megias:2016bde}.

The expected sensitivities at the end of the high luminosity LHC (HL-LHC) run are $\delta {y_f} \sim 5 \%$,
for $f=(t),b,\tau$  (see, {\it e.g.,} \cite{Peskin:2013xra}), 
$\delta \lambda \sim (20-30) \%$ \cite{Goertz:2013kp,Goertz:2014qta,Azatov:2015oxa}, and $\lambda_Z \sim 
10^{-3}$~\cite{Bian:2015zha}\footnote{Note that current experimental constraints are already at the level 
of $\lambda_Z \sim 3 \%$.}, which will set the ballpark for the 
hypothetical measurements considered below. These values could still be improved, for example by the 
ILC, which could allow for $\delta {y_f} \sim 1 \%$, for $f=b,c,\tau$ \cite{Peskin:2013xra},
$\delta \lambda \sim 10 \%$ \cite{Asner:2013psa}, and $\lambda_Z \sim 10^{-4}$ \cite{Bian:2015zha}.
For $C_9$, on the other hand, we consider a value of $C_9 \sim -1$, as suggested by 
experimental anomalies in $B$ physics, see footnote \ref{fn:B}.

In Figs.~\ref{fig:h3-Y}-\ref{fig:c9-TG}, we finally explore the predictions for 
these quantities, induced by non-zero values of the corresponding coefficients, both
for Scenario {\bf A} and {\bf B}. Employing relations (\ref{eq:rel}) and (\ref{eq:rel2}), we can draw 
iso-contours of constant $M$ (dividing regions of different color) in two-dimensional planes 
spanned by the different (pseudo-)observables,
where we can determine the heavy mass $M$ via a combined measurement of the latter.
With a slight abuse of notation, we plot the absolute values of the corresponding quantities,
keeping in mind that the signs of the various coefficients might vary.

We start by studying the correlations between $\delta {y_f}$ and $\delta \lambda$, visualized
in Figure~\ref{fig:h3-Y}. We also present, as colored lines, the values of the coupling $g_\ast$
scanning the planes, where the dot-dashed yellow and orange lines
correspond to NP couplings of $g_\ast = 1,4$, respectively. The green dashed line visualizes the 
boundary of $g_\ast = 8$, beyond which perturbation theory becomes problematic,
while the red line, corresponding to $g_\ast = 4 \pi$, signals the complete breakdown of perturbation
theory. We thus do not draw the colored regions beyond this point.

Looking at the left panel of the figure, representing Scenario {\bf A},
we find that the observation of a deviation in Yukawa couplings of
\begin{itemize}
\item 
$\delta y_f = 15\%$ together with a change in the trilinear Higgs self coupling of $\delta \lambda \sim 8$
indicates NP at $M \approx 3$\,TeV.
\end{itemize} 
Observing on the other hand 
\begin{itemize}
\item
$\delta y_f = 1\%$
and $\delta \lambda \sim 2.5$ leads to a prediction of $M \approx 20$\,TeV,
\end{itemize}
and thus to an explicit sensitivity to the {\it mass}, where a new particle is expected, far beyond
direct collider reach.\footnote{Alternatively, the effect could stem from $N$ particles with
masses of $\sqrt N\, M$.}
The maximally reachable sensitivity (respecting $g_\ast < 8$ and experimental prospects) 
appears for $\delta y_f = 1\%$, $\delta \lambda \sim 3$ and correspond to $M \approx 25$\,TeV.
The corresponding values for all pairs of pseudo-observables for the scenarios at hand 
will be summarized in Table \ref{tab:M}.\footnote{Note that limitations in the 
accuracy of the measurements need to be considered, when determining the different NP masses. 
Nevertheless, the prospective accuracy will for example allow to clearly distinguish the two parameter-space
points considered above, and thus can lead to valuable knowledge on where to expect NP,
see also below.}

Finally, the analysis leads to the further interesting observation that an effect of, say, 
$\sim 20 \%$ in Yukawa couplings {\it requires} in fact sizable deviations in the self coupling of $\sim 250\%$, 
if no new physics resides below $1$\,TeV.
Similarly  
\begin{itemize}
\item$\delta y_f = 40\%$ would require a factor of almost $10$ in 
the self coupling, within LHC reach in the near future.
\end{itemize}
Thus, if such a deviation in yukawa couplings would be observed, while
the self coupling would be constrained to $\delta \lambda <10$ 
(and no new physics would appear below a TeV), the large
class of NP described by Scenario {\bf A} could be basically excluded.
Indeed, for constant $y_f$, 
\begin{itemize}
\item increasing $\delta \lambda$ leads to {\it larger} NP masses,
\end{itemize}
which is due to the peculiar scaling of $\delta \lambda$ with a
large power of $g_\ast$. In that context, the figure finally demonstrates that
sizable deviations in the Higgs trilinear self coupling are possible, 
with small deviations elsewhere and a theory respecting perturbativity, 
as can be seen from the fact that a factor of a few in the former
coupling is consistent with deviations in Yukawa interactions at/below the per cent level
and moderate coupling strength ($g_\ast \ll 4 \pi$), as well as large 
NP masses ($M>10$\,TeV).

The situation is quite different in the SILH-like Scenario {\bf B}, where
perturbativity significantly limits the size of $\delta \lambda$ for moderate
values of $\delta y_f$. A $10\%$ deviation in Yukawa couplings allows at
most for a similar effect in the trilinear coupling, since larger values correspond
to couplings $g_\ast$ entering a terrain where perturbation theory starts to become
unreliable, as depicted by the dashed green line. Measuring, on the other hand, 
simultaneously such deviations,
\begin{itemize}
\item $\delta y_f = \delta \lambda =0.1$, leads to the knowledge that NP 
should show up at $M \approx 8\,$TeV,
\end{itemize}
assuming Scenario {\bf B}.\footnote{This prediction finally needs to be confronted with 
{\it all} pseudo-observables, see below.}
Moreover, seeing 
\begin{itemize}
\item $\delta y_f$ approaching $\sim\!0.5$ and at most  $\delta \lambda \lesssim 0.3$
requires NP appearing below $M \lesssim 3$\,TeV
\end{itemize}
(without very weak couplings, since the orange
dot-dashed line, signaling $g_\ast \leq 4$, does not yet appear) -
otherwise the SILH is not realized in nature.

We now move to Figure~\ref{fig:c9-h3}, where we display the correlation 
between effects in the four-fermion operator ${\cal O}_9$ and $\delta \lambda$.
Beginning again with Scenario {\bf A}, shown in the upper left panel, we observe
a potential sensitivity to very large NP masses, reaching even the $100$\,TeV range. 
As discussed before, in this case there are actual experimental hints for a non-vanishing NP effect,
corresponding to $C_9 \sim -1$, which we depict by the red dashed line.
Motivated by this anomaly, we now exemplify in the lower panel of the Figure in more detail 
a potential determination of the NP mass $M$. Assume that in fact a value of $C_9=-1.2 \pm 0.2$ is established 
in the future~\cite{Altmannshofer:2017fio}, while the trilinear Higgs self-coupling exhibits
a $\delta \lambda = (60 \pm 10) \%$ correction. Given this information, we could conclude that
$M \in \{40,60\}\,$TeV, which is derived from building the minimum and the maximum of $M$ over the 
four corners of the gray box denoted by $\sim 50$\,TeV.
On the contrary, a hypothetical value of  $C_9=-4 \pm 0.2$, together with the constraint $\delta \lambda \lesssim 10 \%$ 
would lead to the prediction $M\lesssim 6$\,TeV, if the underlying framework is Scenario {\bf A}.
The plot in the upper right panel summarizes the results in Scenario {\bf B}. Here, this
pair of observables is less rich, since moderate values of $C_9$ allow at most for $\delta \lambda \lesssim 3\,\%$,
beyond any hope for detectability with current or planned experiments. Still, establishing {\it any} deviation in the triple
Higgs coupling (without an excessive $C_9$) would again exclude the underlying SILH hypothesis.

Next, we consider a simultaneous measurement of the TGC parameter $\lambda_Z$ and $\delta y_f$ in Scenario {\bf A}. 
Here, a pretty strong constraint/exact measurement of a potential deviation in $\lambda_Z$ is necessary, 
together with non-vanishing effects in $\delta y_f$, approaching the perturbativity bound, in order to reach 
high masses. For example, $\lambda_Z \lesssim 1.5\,\%$ and $\delta y_f \sim 20\,\%$ leads to 
$M\approx 5\,$TeV, as can be read off from the upper left panel of Figure~\ref{fig:Y-TG},
while $\lambda_Z \lesssim 10^{-3}$ and $\delta y_f \sim 1\,\%$ leads to $M \approx 20\,$TeV,
as is visible from the zoom into the small $\lambda_Z$ region in the lower left panel. 
In Scenario {\bf B}, the scaling of $M$ with $\lambda_Z$ is inverted, due to the additional loop
factor entering $c_{3V}$. While a $\delta y_f = 5\,\%$ deviation in Yukawa couplings together with 
a $\lambda_Z = 5 \times 10^{-4}$ effect features $M \approx 3\,$TeV, the same deviation coming
with  $\lambda_Z = 1.5 \times 10^{-3}$ leads to $M \approx 10\,$TeV. 
Large effects in $\lambda_Z$ together with small $\delta y_f$, such as $\lambda_Z \gtrsim 1\,\%$ 
and $\delta y_f \lesssim 30\,\%$ are not possible in the perturbative regime, while the inverted
case of $\lambda_Z \lesssim 2 \times 10^{-3}$ and $\delta y_f \gtrsim 30\,\%$ is in conflict
with the non-observation of new sub-TeV particles. In fact, sizable effects in one of the pseudo-observables
require also non-negligible effects in the other. The same tendency holds also in Scenario {\bf A}.
In summary, while for the pair of $(\lambda_Z,\delta y_f)$ the potential to determine large NP masses
is not tremendous, the strong correlations offer a valuable means to test/rule out the SILH or ALH paradigms.
Moreover, the observation that the sensitivity to heavy masses increases if the coefficient that features the smaller 
power of $g_\ast$ becomes stronger constrained (which is $\lambda_Z$ ($\delta y_f$) in Scenario {\bf A} ({\bf B})) is nicely confirmed in the plots.

Another interesting pair of (pseudo-)observables is $\delta \lambda$ and $\lambda_Z$, explored in 
Figure~\ref{fig:h3-TG}. In Scenario {\bf A}, given in the left panel of the figure,
large corrections to the trilinear Higgs self coupling are viable, with no 
measurable impact on $\lambda_Z$. Observing for example a $\delta \lambda = 100\,\%$ correction
simultaneously with $\lambda_Z = 2 \times 10^{-4}$ is consistent with the framework and corresponds to
a very large NP mass of $M=50\,$TeV! Similarly, $\lambda_Z = 2 \times 10^{-3}$ allows for 
$\delta \lambda = 8$ $-$ a measurement which would indicate that $M\approx15\,$TeV.
In Setup {\bf B}, on the other hand, $\delta \lambda = 100\,\%$ requires $\lambda_Z \gtrsim 3\,\%$ 
(see right panel of the figure), which is already in tension with current limits.
NP masses corresponding to combined experimental results in this plane within collider reach are rather low, 
not exceeding $M\approx 8\,$TeV, which is reached for $\delta \lambda = 10\,\%$ and $\lambda_Z=3 \times 10^{-3}$.

The remaining plane to be discussed is spanned by $C_9$ and $\lambda_Z$. Here we can focus on Scenario {\bf A},
depicted in the left panel of Figure~\ref{fig:c9-TG}, since measurable effects in Scenario 
{\bf B} (in $\lambda_Z$), explored in the right panel of the same figure, are basically excluded from perturbativity.
In the former setup, however, largely different NP masses can be discriminated. While $\lambda_Z = 0.4\,\%$
with $C_9=-1$ leads to the prediction $M \approx 2\,$TeV, just at the boundary of the current direct reach,
extracting $\lambda_Z = 0.1\,\%$ and $C_9=-1.5$ induces $M\approx 10\,$TeV, and 
$\lambda_Z = 3 \times 10^{-4}$ with the same value for $C_9$ allows the conclusion that
$M \approx 35\,$ TeV. At the same time, Scenario {\bf B} strictly predicts a very small $\lambda_Z \sim {\cal O}(10^{-4})$,
given a moderate $C_9$ not exceeding ${\cal O}(1)$. 
While for $\lambda_Z = 10^{-4}$ and $C_9=-1.5$ one finds $M \approx 50\,$TeV, a 
sizable $\lambda_Z$ basically excludes this scenario.

Before concluding we will finally discuss the procedure lined out below eq. (\ref{eq:rel}) for two simple explicit examples.
To this end, consider first a measurement of $\lambda_Z \approx 0.15\,\%$ and $\delta \lambda \approx 10\,\%$.\footnote{
Although this is a far-future scenario, it serves the illustrative purpose.}
Employing Figure~\ref{fig:h3-TG} (or eq. (\ref{eq:rel})), we deduce $M\approx 18\,$TeV for Scenario~{\bf B} and 
$M\approx 9\,$TeV for Scenario {\bf A}, with couplings $g_\ast \approx  10$ and $g_\ast \approx 2$, respectively.
With this information, we derive via eqs.~(\ref{eq:sys}) and (\ref{eq:rel}) $\delta y_f \approx 3\,\%$ with $C_9 \approx -18$
in Scenario~{\bf B}, while in Scenario~{\bf A} we obtain $\delta y_f \approx 0.5\,\%$ and $C_9 \approx -2$.
While Scenario {\bf B} is clearly excluded, establishing a $C_9 \lesssim -(1-2)$ would lead to a consistent picture
of effects in Scenario {\bf A}. Thus, given that nature respects approximately the scaling lined out in Table \ref{tab:cor2}, 
we could conclude that NP is present at $M \approx 10\,$TeV, with a moderate coupling strength $g_\ast \approx 2$, 
in the reach of a future collider.

Clearly, less involved examples exist, where directly one of the two orthogonal scenarios is excluded.
Consider thus finally $\delta \lambda = 70\,\%$ and $C_9 = -1.2$. From Figure~\ref{fig:c9-h3} we can directly exclude
Scenario {\bf B}. For Scenario {\bf A}, on the other hand, we derive $M\approx 50\,$TeV and $g_\ast \approx 8$.
This leads now to the predictions $\delta y_f \approx 0.2\,\%$ and $\lambda_Z \approx 2 \times 10^{-4}$, obviously
in agreement with current limits, such that we obtain a consistent picture of rather strongly coupled NP
significantly above energies testable at (near) future colliders.

We conclude this section noting that in less simple concrete models the final Wilson coefficients might deviate by a numerical ${\cal O}(1)$ factor 
from the above estimates - however, given that the scenario is broadly characterized by a single relevant NP coupling strength and a single mass scales
(and not conflicting the broad assumptions we detailed), the NP mass that the analysis at hand will point to stays in the same 
ballpark and general correlations and tendencies remain valid.


\section{Conclusions}
\label{sec:conc}

We have shown how simultaneous measurements of (pseudo-)observables allow to determine
explicitly the mass of NP particles, beyond direct reach. This is achieved by exploiting their different 
scaling with the NP coupling $g_\ast$, as derived after restoring the $\hbar$ dimensions of the operators, 
with the details of the procedure worked out in the body of the paper.

To summarize the results from the perspective of {\it expected} effects in the two orthogonal scenarios considered,
in the ALH-like Scenario {\bf A} ${\cal O}(1)$ effects are possible in $\delta \lambda$, without other
problematic contributions, while sizable $\delta y_f$ of $\gtrsim 10\,\%$ require a large $\lambda_Z \gtrsim 1\,\%$ 
and a $C_9$ exceeding significantly experimental limits (under the given flavor hypothesis). 
Even a $\delta y_f \gtrsim 1\,\%$ leads to $\lambda_Z \gtrsim 0.1\,\%$ and $|C_9| \gtrsim 5$.
In turn, sizable $C_9$ are possible without inducing large corrections to other pseudo-observables.
Finally, a large $\lambda_Z$ at the per cent level induces also a detectable $\delta y_f$, unless the
mass of the new physics is very low ($M \lesssim 1$\,TeV), along with $|C_9| \gtrsim {\cal O}(1)$, 
while $\delta \lambda$ can easily remain at an unobservable level.

On the other hand, the SILH-like Scenario {\bf B} predicts basically tiny $\delta \lambda$
and $\lambda_Z$, below a detectable level, since otherwise $|C_9|$ becomes too large
(requiring perturbativity).
Even ignoring the flavor-structure related $C_9$, sizable $\delta \lambda$ and 
$\lambda_Z$ at a level detectable {\it in the near future} are disfavored, since they also 
induce large corrections in Yukawa couplings around the current experimental
sensitivity. Finding however a sizable $\delta y_f \sim 25 \%$ would in turn
lead unavoidably to a $\lambda_Z$ observable in the long-term LHC run
(requiring $M\gtrsim 1\,$ TeV) and to a $C_9$ vastly exceeding limits. 
The strong and correlated predictions in this scenario offer a powerful means to
test it {\it indirectly} in the near future with simple observations.
Finally, sizable $C_9$ are viable in Scenario {\bf B}, without any other observable prediction.
The $(\delta \lambda, C_9)$ plane is particularly promising for testing large $M$ in Scenario {\bf A},
while in Scenario {\bf B} the same holds true for $(\lambda_Z,C_9)$ and the
figures can also be used to estimate if NP is expected to be detected first directly or indirectly.

The presented approach allows to test simple NP frameworks, 
such as the SILH or ALH, and to arrive at a mass $M$, as a (unique) solution to the experimental picture.
While the article lines out the general procedure, in the presence of a signal
dedicated studies along the lines worked out above
will be in order to finally unveil the nature of the NP.
After all, this work provides on overview of which patterns of NP can be expected 
and how an indication for the NP mass can be obtained, given only indirect observation,
which can help to find the UV completion of the SM and to develop strategies to detect the NP {\it directly}.

\begin{table}[t]
\centering
\begin{tabular}{|c||c|c|c|c|}
\hline
 {\bf A} & $\delta {y_f}$ & $\delta\lambda$ & $\lambda_Z$ & $c_9$ \\
\hline
\hline
$\delta {y_f}$ & - & \small 25\,TeV& \small 25\,TeV & - \\
\hline
$\delta\lambda$ &  \hspace{-2mm} \small 25\,TeV \hspace{-2mm} & - & \small 70\,TeV&  \hspace{-2mm} \small 150\,TeV \hspace*{-2mm}\\ 
\hline
$\lambda_Z$ & \hspace{-2mm}  \small 25\,TeV \hspace{-2mm} & \small 70\,TeV & - & \hspace{-2mm}\small 70\,TeV\hspace*{-2mm}\\
\hline
$C_9$ & - & \hspace{-2mm} \small 150\,TeV \hspace{-2mm} & \small 70\,TeV & - \\
\hline
\end{tabular}
\quad
\begin{tabular}{|c||c|c|c|c|}
\hline
 {\bf B} & $\delta {y_f}$ & $\delta\lambda$ & $\lambda_Z$ & $c_9$ \\
\hline
\hline
$\delta {y_f}$ & - &  \small 8\,TeV & \small 25\,TeV & - \\
\hline
$\delta\lambda$ &  \small 8\,TeV & - & \small 8\,TeV &  \small  N/A\\ 
\hline
$\lambda_Z$ & \hspace{-2mm} \small 25\,TeV \hspace{-2mm} 
& \small 8\,TeV & - & \hspace{-1.mm}\small 40\,TeV\hspace*{-1.mm}\\
\hline
$C_9$ & - &  \small  N/A  &\small 40\,TeV & - \\
\hline
\end{tabular}
\caption{\label{tab:M} Maximal mass $M$ detectable, for each pair of pseudo-observables, considering
ILC projections, see text for details.}
\end{table}

\section*{Acknowledgments}
I am indebted to Roberto Contino, Adam Falkowski,  Christophe Grojean, and Francesco Riva
for valuable discussions on EFT extensions of the SM and to Adam Falkowski and Francesco
Riva for useful remarks on the manuscript. Finally, I thank the CERN theory division for the 
wonderful time during my fellowship (2014-16), when most of this work was performed.

\end{document}